\documentclass[a4paper,fleqn,usenatbib,useAMS]{mnras}

\usepackage{graphicx}	
\usepackage{amsmath}	
\usepackage{amssymb}	
\usepackage{multicol}        
\usepackage{bm}		
\usepackage{pdflscape}	

\usepackage[T1]{fontenc}
\usepackage{ae,aecompl}

\usepackage{mathptmx}

\title[Resolved diagnostic diagrams: LIER galaxies] 
{SDSS IV MaNGA - Spatially resolved diagnostic diagrams: \\
A proof that many galaxies are LIERs}
\author[F. Belfiore et al.] 
{Francesco Belfiore$^{1,2}$\thanks{Email: fb338@cam.ac.uk},
Roberto Maiolino$^{1,2}$,
Claudia Maraston$^{3}$,
Eric Emsellem$^{4,5}$,
\newauthor 
Matthew A. Bershady$^{6}$, 
Karen L. Masters$^{3, 7}$ 
Renbin Yan$^{8}$,
Dmitry Bizyaev$^{9, 10}$,
\newauthor 
M\'ed\'eric Boquien$^{11}$,
Joel R. Brownstein$^{12}$, 
Kevin Bundy$^{13}$,
Niv Drory$^{14}$,
\newauthor 
Timothy M. Heckman$^{15}$,
David R. Law$^{16}$,
Alexandre Roman-Lopes$^{17}$,
Kaike Pan$^{9}$,
\newauthor 
Letizia Stanghellini$^{18}$,
Daniel Thomas$^{3}$, 
Anne-Marie Weijmans$^{19}$ and 
Kyle B. Westfall$^{3}$.
\\
\\
(Affiliations can be found after the references)
}

\pagerange{\pageref{firstpage}--\pageref{lastpage}} 
\pubyear{2016}

\begin{document}
\label{firstpage}

\maketitle

\begin{abstract}
We study the spatially resolved excitation properties of the ionised gas in a sample of 646 galaxies using integral field spectroscopy data from SDSS-IV MaNGA. Making use of Baldwin-Philips-Terlevich diagnostic diagrams we demonstrate the ubiquitous presence of extended (kpc scale) low ionisation emission-line regions (LIERs) in both star forming and quiescent galaxies. In star forming galaxies LIER emission can be associated with diffuse ionised gas, most evident as extra-planar emission in edge-on systems. In addition, we identify two main classes of galaxies displaying LIER emission: `central LIER' (cLIER) galaxies, where central LIER emission is spatially extended, but accompanied by star formation at larger galactocentric distances, and `extended LIER' (eLIER) galaxies, where LIER emission is extended throughout the whole galaxy. 

In eLIER and cLIER galaxies, LIER emission is associated with radially flat, low H$\alpha$ equivalent width of line emission ($<$ 3 \AA) and stellar population indices demonstrating the lack of young stellar populations, implying that line emission follows tightly the continuum due to the underlying old stellar population. The H$\alpha$ surface brightness radial profiles are always shallower than $\rm 1/r^{2}$ and the line ratio [OIII]$\lambda$5007/[OII]$\lambda$3727,29 (a tracer of the ionisation parameter of the gas) shows a flat gradient. This combined evidence strongly supports the scenario in which LIER emission is not due to a central point source but to diffuse stellar sources, the most likely candidates being hot, evolved (post-asymptotic giant branch) stars. 
Shocks are observed to play a significant role in the ionisation of the gas only in rare merging and interacting systems.

\end{abstract}


\section{Introduction} 
\label{intro}

The properties of galaxies in terms of their star formation activity are observed to follow a bimodal distribution, often expressed in terms of a `red sequence' and `blue cloud' \citep{Strateva2001, Blanton2003, Baldry2004, Baldry2006}. These two sequences are clearly separated not only in their integrated optical and UV colours \citep{Wyder2007}, but also in their stellar populations \citep[e.g][]{Kauffmann2003b}, cold gas content and ionised gas properties \citep{Kewley2006, Saintonge2011}.

Galaxies in the blue cloud are characterised by an emission line spectrum typical of classical H\textsc{ii} regions, with nebular line ratios varying mostly as a function of metallicity and ionisation parameter \citep{Kewley2001, Levesque2010}. Conversely, galaxies on the red sequence host only low levels of residual line emission \citep[e.g.][]{Phillips1986, Goudfrooij1999, Sarzi2010} with line ratios generally inconsistent with those expected from star formation. Their spectra are characterised by strong low-ionisation transitions (e.g. [OI]$\lambda$6300, [SII]$\lambda\lambda$6717,31) and display the characteristic line ratios of `low ionisation nuclear emission line regions' (LINERs, \citealt{Heckman1980}). 

Originally associated with weak active galactic nuclei \citep{Kauffmann2003a, Kewley2006, Ho2008}, LINER emission has lately attracted increasing amounts of attention. Spatially resolved observations have demonstrated that LINER-like emission is not confined to nuclear regions in galaxies, but often appears to be extended on kpc scales \citep{Sarzi2006, Sarzi2010, Singh2013}, especially in early type galaxies. In these galaxies, several authors have argued that the extended LINER-like emission is consistent with photoionisation by hot evolved stars (and in particular post asymptotic giant branch stars), which become the main source of ionising photons in quiescent regions after star formation has ceased \citep{Binette1994, Stasinska2008, CidFernandes2011, Yan2012}. While observational support for the `stellar hypothesis' for LINER emisson is mounting, previous work has been limited by lack of spatial resolution \citep[e.g.][]{Yan2012}, wavelength coverage \citep[e.g.][]{Sarzi2010} or small sample size \citep[e.g.][]{Singh2013}.

In this work we make use of a large sample of 646 galaxies of all morphological types and covering a wide range in stellar mass ($\rm 10^9 - 10^{12} M_\odot$) observed with integral field spectroscopy (IFS) to map the excitation properties of the ionised gas and study the occurrence and properties of LINER-like emission. Our data comes from the Sloan Digital Sky Survey-IV (SDSS-IV) Mapping Nearby Galaxies at Apache Point Observatory (MaNGA) survey \citep{Bundy2015}, an IFS survey targeting a statistically representative sample of 10~000 nearby ($\rm <z> \sim 0.03$) galaxies. Crucially for the aim of this work, MaNGA's observed wavelength range (3600 -10300 \AA) covers all the strong nebular lines used in the standard BPT \citep{Baldwin1981, Veilleux1987} excitation diagnostic diagram. Moreover, the large and well-defined sample, combined with the uniform spatial coverage of MaNGA (covering galaxies out to at least 1.5 effective radii [$\rm R_e$]), allows us to study ionised gas across both the red sequence and the blue cloud at kpc spatial resolution.

In this paper we focus on the spatially resolved observables that are most relevant for shedding new light onto the excitation source of LINER-like emission. In a companion paper (Belfiore et. al., \textit{in prep.}, henceforth \textit{Paper II}) we focus on the global properties of LINER-like galaxies, with the aim of placing this class of galaxies in context within the bimodal galaxy population.

This work is structured as follows. In Sec. \ref{sec2} we describe the MaNGA dataset and our customised spectral fitting procedure. In Sec. \ref{sec3} we review the standard BPT diagram and predictions from ionisation from hot evolved stars. In Sec. \ref{sec4} we populate the BPT diagram with the (kpc-scale) resolved MaNGA data. Based on the distribution of star forming regions and LINER-like emission within galaxies (excitation diagnostic maps) we then propose a new classification scheme that divides LINER-like galaxies in two classes, depending on the presence or absence of star formation at large radii. In the following sections we study the properties of the ionised gas through the H$\alpha$ emission (Sec. \ref{sec6}), diagnostic line ratios (Sec. \ref{sec7}) and age-sensitive stellar population indices (Sec. \ref{sec8}) to test the `stellar hypothesis' for LINER-like emission. We discuss the strengths and weaknesses of the stellar hypothesis for LINER-like emission in Sec. \ref{dis} and conclude in Sec. \ref{sum}.

Throughout this work redshifts, photometry and stellar masses are taken from a customised version of the Nasa Sloan Atlas (NSA\footnote{http://www.nsatlas.org}, \citealt{Blanton2005, Blanton2005a, Blanton2011}). Stellar masses are derived from SDSS photometry using the {\tt kcorrect} software package (version {\tt v4\_2}, \citealt{Blanton2007}) with \cite{Bruzual2003} simple stellar population models and assuming a \cite{Chabrier2003} initial mass function. Effective radii are measured from the SDSS photometry by performing a S\'ersic fit in the r-band. Single fibre spectroscopic data from the SDSS Legacy Survey data release 7, \citep{Abazajian2009} for the main galaxy sample targets \citep{Strauss2002} are referred to as `legacy SDSS'.
When quoting luminosities, masses and distances we make use of a $\Lambda$CDM cosmology with $\Omega=0.3$, $\Lambda=0.7$ and $\rm H_0=70 \mathrm{ \ km^{-1} s^{-1}Mpc^{-1}}$.

\section{The MaNGA data}
\label{sec2}

\subsection{Overview of MaNGA observations}
\label{sec2.1}

The MaNGA sample is constructed to target local ($\rm 0.01< z < 0.15 $) galaxies and meet two fundamental design criteria: a flat distribution in i-band absolute magnitude ($\rm M_i$, as a proxy for stellar mass) and uniform radial coverage in terms of galaxy $\rm R_e$. To achieve both goals, the MaNGA instrument consists of a set of 17 hexagonal integral field units (IFUs) of different sizes, ranging from 19 (12$''$ on sky diameter) to 127 (32$''$ on sky diameter) fibres. 12 additional mini-bundles (7 fibres each) are used to observe standard stars for flux calibration and 92 single fibres are used for sky subtraction. All fibres have an on sky diameter of 2$''$ and feed the dual beam BOSS spectrographs \citep{Smee2013}, which cover continuously the wavelength range between 3600 \AA\ and 10300 \AA\ with a spectral resolution varying from R $\sim$ 1400 at 4000 \AA\ to R $\sim$ 2600 at 9000 \AA. The MaNGA instrument suite is mounted on the SDSS 2.5m telescope at APO \citep{Gunn2006} and is described in \cite{Drory2015}

\begin{figure} 
\includegraphics[width=0.50\textwidth, trim=40 0 10 60, clip]{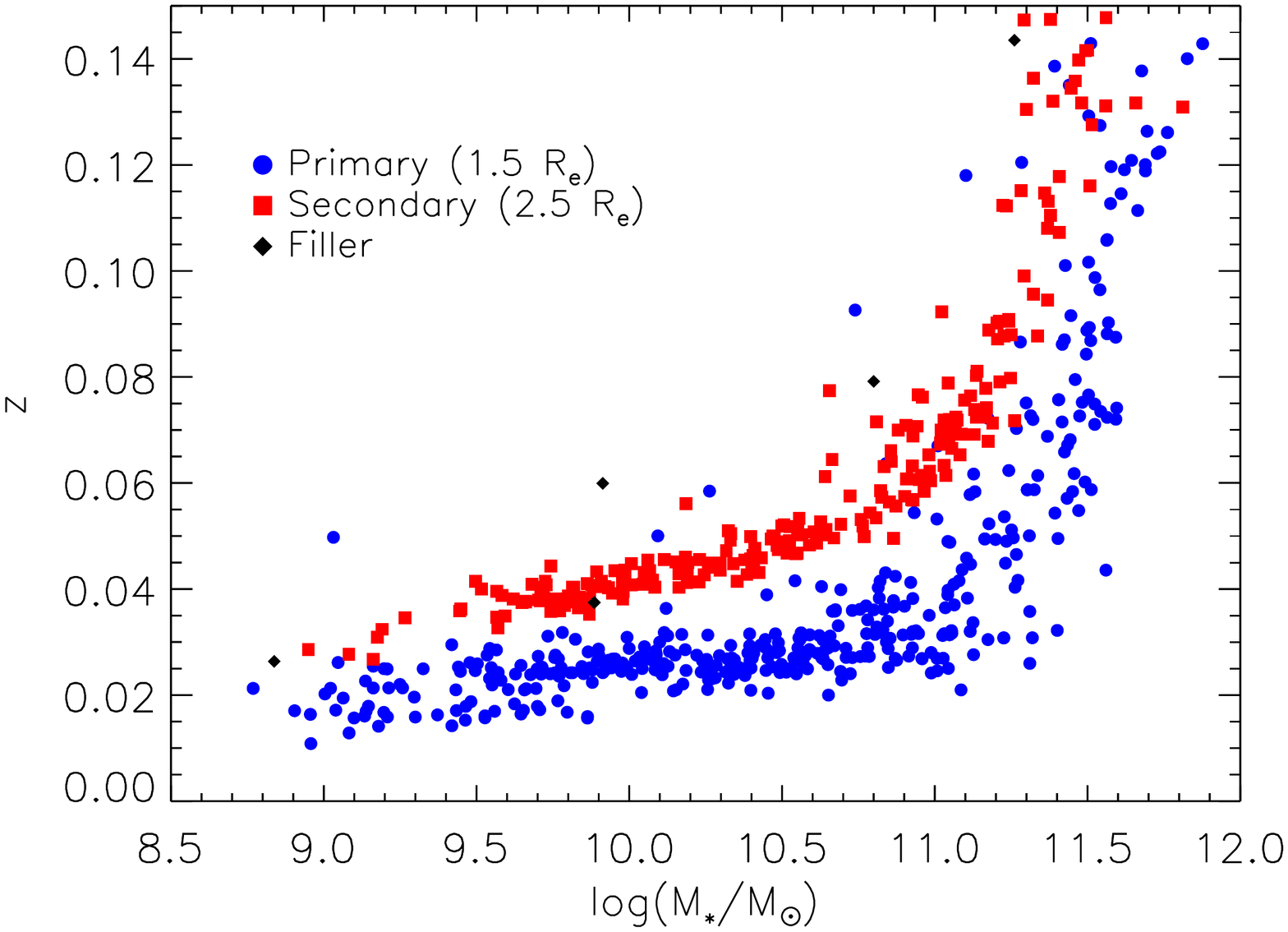}
\includegraphics[width=0.50\textwidth, trim=40 10 0 50, clip]{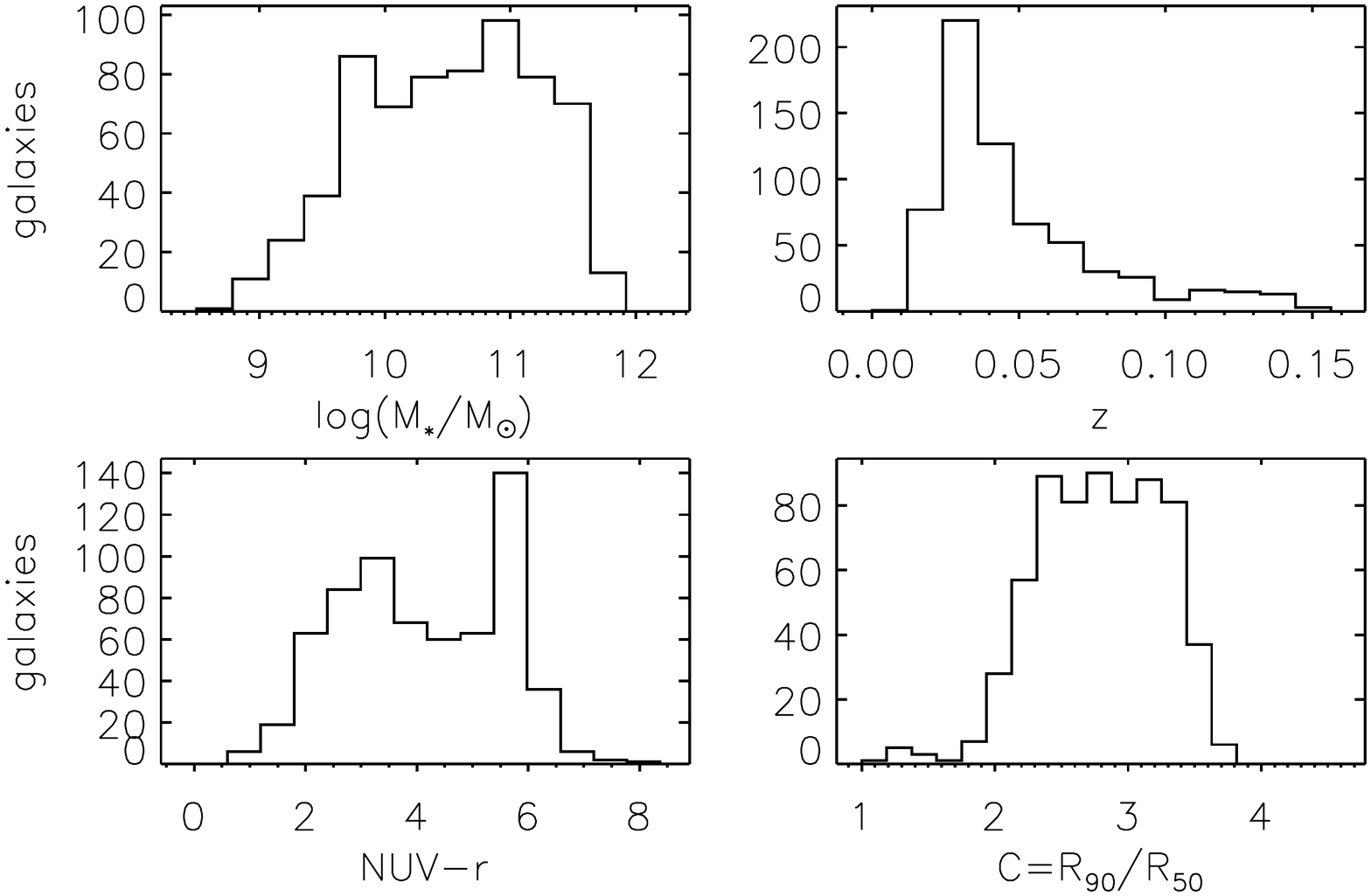}
\caption{Top: The MaNGA sample used in this work in the stellar mass-redshift plane. Due to the fixed size distribution of the MaNGA bundles and the fact that MaNGA observes galaxies out to a fixed galoctocentric distance in terms of $\rm R_e$, higher mass galaxies are preferentially observed at higher redshift. For an equivalent reason the secondary sample (radial coverage out to 2.5 $\rm R_e$) is preferentially observed at higher redshift at fixed stellar mass. A handful of galaxies have been observed while not fitting in either the primary or secondary sample selection functions and are designed as `filler'. Bottom: Histogram distributions of stellar mass, redshift, $NUV - r$ colour and concentration parameter ($\rm C = R_{90}/R_{50}$, where R are Petrosian radii) for the galaxy sample used in this work. We note that our sample is roughly flat in stellar mass in the range $\rm \log(M/M_{\star}) >9.0$, has median redshift  $\rm <z> \sim 0.037$ and samples well both the blue cloud and the red sequence.}
\label{fig2.1}
\end{figure}

MaNGA target galaxies are divided in a `primary' and a `secondary' sample, following a 3:1 ratio. For primary sample galaxies, MaNGA aims to obtain radial coverage out to at least 1.5 $\rm R_e$, while for secondary sample galaxies MaNGA observes out to at least 2.5 $\rm R_e$. The mean redshift of the full MaNGA sample is $\rm <z> \sim 0.03$, with secondary sample galaxies being generally at higher redshift ($\rm <z> \sim 0.045$). For each bin of $\rm M_i$ the MaNGA target selection defines a minimum and maximum redshift for observations, within which the MaNGA sample is volume limited. Further details on the MaNGA sample selection and bundle size optimisation are presented in Wake et al. ({\it in prep.}).
The stellar mass and redshift distribution of the sample used for this work is shown in Fig. \ref{fig2.1}. Our sample includes all MaNGA galaxies observed before April 2015 (first 8 months of survey operations) and consists of 646 unique galaxies (not taking into account the possible presence of smaller companions/satellites within a single MaNGA bundle). The properties of this sample compare well with the properties of the full MaNGA parent sample in terms of stellar mass coverage (approximately flat in stellar mass for $\rm \log(M_\star/M_{\odot}) >9$) and redshift ($<z> \sim 0.037$).

On average each MaNGA plate is exposed until it reaches a fixed target S/N level. A three point dithering pattern is applied to compensate for light loss between fibres and to obtain a uniform circular point spread function (PSF, \citealt{Law2015}). The MaNGA data was reduced using version {\tt v1\_3\_3} of the MaNGA reduction pipeline (Law et al., {\it submitted}). The relative spectrophotometry is found to be accurate to within 1.7$\%$ between H$\alpha$ and H$\beta$ and to within 4.7$\%$ between [NII]$\lambda6584$ and [OII]$\lambda\lambda 3726,29$ \citep{Yan2016}. The wavelength calibrated, sky subtracted and flux calibrated MaNGA fibre spectra (error vectors and mask vectors) and their respective astrometric solutions are combined to produce final datacubes with pixel size set to 0.5$''$. The median PSF of the MaNGA datacubes is estimated to have a full width at half maximum (FHWM) of 2.5$''$. Since every pixel is associated with spectral information, for the rest of this work we refer to the MaNGA pixels in the spatial dimension as \textit{spaxels}.

\subsection{Spectral fitting}
\label{sec2.2}

We extract kinematics, emission line fluxes and stellar population indices from the MaNGA data through a customised spectral fitting procedure, which models both the stellar continuum and the nebular emission lines. Full details of the spectral fitting procedure and associated uncertainties, together with further details of the MaNGA data analysis pipeline, will be presented in future papers in this series. In the following we give a brief summary of the basic steps.

\begin{enumerate}
\item{We perform an initial Voronoi binning \citep{Cappellari2003} of the MaNGA datacubes based on the continuum signal to noise (S/N) in the 6000-6200 \AA\ range. We choose a S/N target of 6 (per 69 $\mathrm{ km \ s^{-1}}$ channel), which is well suited to extracting reliable kinematic from the continuum (but not sufficient to deriving higher Gauss-Hermite moments of the the stellar line of sight velocity distribution). We do not include in the binning procedure spaxels with S/N $<$1. The effect of noise covariance between spaxels is taken into account by modifying the original Voronoi binning procedure of \cite{Cappellari2003} and scaling the binned noise vector according to an analytical prescription similar to the one adopted by the CALIFA survey \citep{Garcia-Benito2015}.}

\item{The resulting binned spectra are fitted with a set of simple stellar population (SSP) models using penalised pixel fitting (PPXF, \citealt{Cappellari2004}). A window of 1200 $\rm km \ s^{-1}$ around the expected position of emission lines (taking into account the galaxy redshift) is excluded from the fit. Strong sky lines are also masked. A set of 36 templates from the MIUSCAT SSP library \citep{Vazdekis2012}, spanning a wide range in stellar age (from 60 Myr to 15 Gyr) and four metallicities ([Z/H]=0.2, 0.0, -0.4, -0.7) are used as a basis set for the fitting. For each spaxel we fit for a Gaussian line of sight velocity distribution (and thus derive the velocity [$v_\star$] and velocity dispersion [$\sigma_\star$] of the stellar component) together with dust extinction (using a \citealt{Calzetti2000} extinction law) and a set of additive polynomials up to the 4th order, to take into account possible residual imperfections in the relative spectrophotometry.}

\item{Since the binning on the continuum S/N is not ideal for studying the emission lines, the best fit continuum within each bin is scaled to the flux in the 6000-6200 \AA\ range and subtracted from the observed spectrum spaxel-by-spaxel. An emission-line-only datacube is thus obtained on a spaxel-by-spaxel basis. After this step, a second Voronoi binning based on the S/N around H$\alpha$ is performed. In doing so we allow the inclusion of spaxels that had been previously rejected by the continuum-based Voronoi binning for having undetected continuum. This approach allows us to obtain independent binning schemes for the study of nebular lines and stellar continuum, which is highly advantageous since line and continuum emission can have different surface brightness profiles.}

\item{For each bin the emission lines are fitted with a set of Gaussians, one per line. To increase the ability to fit weaker lines the velocities of all lines are tied together, thus effectively using the stronger lines to constrain the kinematics of the weaker ones. In the case of the [OIII]$\lambda\lambda$4959,5007 and [NII]$\lambda\lambda$6548,83 doublets their dispersions are tied together and their amplitude ratios are fixed to the ratios of the relative Einstein coefficients. Fluxes are obtained by integrating the Gaussian best fit to the line. Errors on the flux are derived from the formal errors on the amplitude and dispersion of the Gaussian fit.}
\end{enumerate}

Based on Monte Carlo simulations of model data to which a suitable noise model was added and analysis of repeat observations, we conclude that, when an emission line is detected with S/N $>3$, the uncertainties calculated by our pipeline are realistic under a wide range of conditions\footnote{In case of a spectral line S/N is defined as the \textit{amplitude} of the fitted Gaussian divided by the average noise in a nearby bandpass.}. The uncertainties are less reliable for galaxies with stronger continuum and older stellar populations, for which an accurate continuum subtraction plays a key role. Simulations show that our spectral fitting procedure is able to recover line fluxes with negligible bias and less than 0.12 dex error for S/N $>$ 3 for all strong lines except H$\beta$, which suffers from larger errors ($\sim 0.2$ dex) and a small positive bias (generally less than $0.08$ dex), possibly due to the uncertainties associated with the continuum subtraction. 

In this work we measure the equivalent width of the H$\alpha$ line (EW(H$\alpha$)) after having subtracted the Balmer absorption due to the best fit stellar continuum, i.e. we consider only the \textit{emission component} of H$\alpha$. For simplicity emission line EWs are defined to be positive in emission.


\section{Emission line diagnostic diagrams and the role of LI(N)ER excitation} 
\label{sec3}

\begin{figure*} 
\includegraphics[width=0.49\textwidth, trim=70 150 90 200, clip]{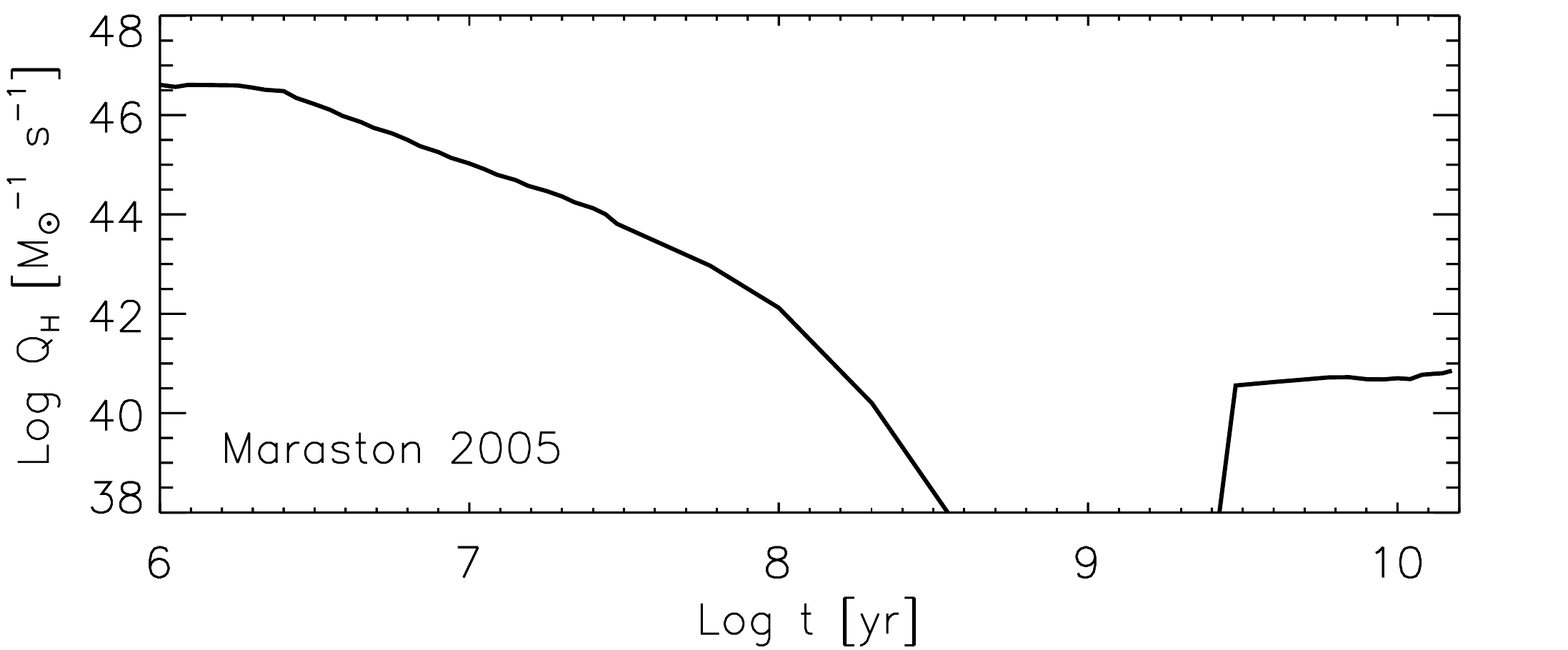}
\includegraphics[width=0.49\textwidth, trim=70 150 90 200, clip]{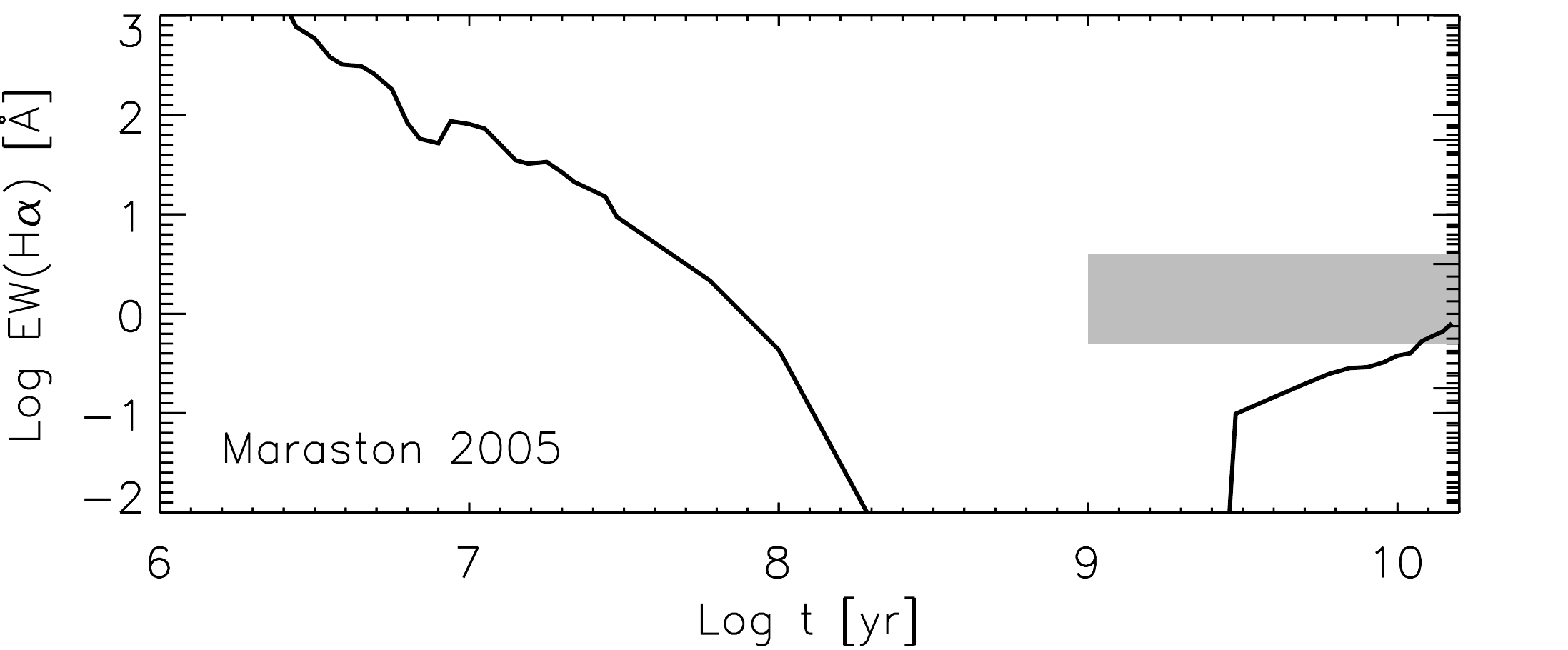}

\caption{The ionising photon flux ($\rm Q_H$) per unit stellar mass and the equivalent width of H$\alpha$ as a function of age after the starburst for \protect\cite{Maraston2005} stellar population models, using pAGB (including planetary nebulae) model tracks based on \protect\cite{Stanghellini2000}. The ionising flux and $\rm EW(H\alpha)$ due to the young stellar population declines quickly after the burst. Physically motivated choice of pAGB parameters predict an $\rm EW(H\alpha)$ in the range 0.5 - 3.0 \AA\ at late times, corresponding to the shaded in grey in the right panel.} 
\label{fig3.2}
\end{figure*}

\subsection{The classical star formation-AGN division}
\label{Sec3.1}

The most widely used system for spectral classification of emission line galaxies is based on the diagnostic diagrams originally suggested by \cite{Baldwin1981} and \cite{Veilleux1987} (generally referred to as BPT diagrams). 
Standard BPT diagnostic diagrams rely on the line ratios [OIII]$\lambda 5007$/H$\beta$ versus [NII]$\lambda 6583$/H$\alpha$, [SII]$\lambda 6717,31$/H$\alpha$ or [OI]$\lambda 6100$/H$\alpha$\footnote{From now on these lines will be referred to as just [OIII], [NII] and [SII].}. This scheme has two key strengths: in $\rm z \sim 0$ Universe all the required lines lie in the easily accessible optical range and the vicinity in wavelength of the pairs of lines considered makes the scheme reddening insensitive.

With the help of large spectroscopic surveys, emission line galaxies have been shown to lie on two distinct sequences in the BPT diagrams \citep{Kauffmann2003a}. The `left-hand' branch (so-called because it is positioned on the left in the traditional BPT diagrams, see Fig. \ref{fig3.3}) is associated with H\textsc{ii} (star forming) regions. The overall shape of this branch can be reproduced well by H\textsc{ii} regions photoionisation models as the envelope of the expected line ratios for different values of metallicity and ionisation parameter for gas ionised by OB stars \citep{Kewley2001, Nagao2006, Stasinska2006, Sanchez2015}. 
The `right-hand' branch in the BPT diagnostic diagrams is generally associated with two populations of sources: high ionisation Seyfert (Sy) galaxies and low ionisation (nuclear) emission line regions [LI(N)ERs].

\cite{Heckman1980} first presented a detailed study of LI(N)ER emission, arguing that LI(N)ERs could represent the low-luminosity extension of the Sy population. However, as already discussed in \cite{Heckman1980}, both shock ionisation models \citep{Dopita1995, Dopita2015} and photoionisation by a hard radiation field \citep{Ferland1983, Kewley2006} can be invoked to explain LINER emission. In more recent years, models for LI(N)ER-like emission due to shock excitation have been further applied outside the AGN framework to merging systems and ultra luminous infrared galaxies (ULIRGs, \citealt{Rich2010, Rich2013}).

 \subsection{Hot evolved stars as the source for LI(N)ER emission}
 \label{sec3.2}
 
Aiming to model extended LI(N)ER-like ionised gas emission in elliptical galaxies, \cite{Binette1994} studied the ionising spectrum due to post-asymptotic giant branch (pAGB) stars predicted by the stellar population models of \cite{Bruzual1993}. pAGB stars in this context include all the stages of stellar evolution subsequent to asymptotic giant branch, including the hydrogen burning, white dwarf cooling and intermediate phases (see \citealt{Stanghellini2000} for a detailed discussion). Since the time spent by stars in the pAGB phase is a strong function of the core mass, only stars within a rather small mass range (core mass 0.55-0.64 $\rm M_\odot$, or initial mass 1-3 $\rm M_\odot$) appear as planetary nebulae \citep{Tylenda1989, Weidemann2000, Buzzoni2006}. In the case of pAGB stars with smaller core masses, the hydrogen ionisation temperature is reached too slowly and only after the material expelled during the AGB phase has already been lost to the general ISM. These stars have been referred to as `lazy pAGB' stars. pAGB stars are generally hot, with typical temperatures around 30~000 K, and are thus ideal candidates as sources of ionising and UV photons. Indeed, stellar population models demonstrate that these stars are the main source of the ionising photon background in galaxies once star formation has ceased \citep{Binette1994}.

The low H$\alpha$ equivalent widths (EW(H$\rm\alpha) < 3~ \AA$) observed in most early type galaxies are compatible with the expectations from reprocessing of the ionising radiation from hot evolved stars \citep{Binette1994, Stasinska2008, CidFernandes2011}, although some controversy persists in the literature as to whether equivalent widths larger than 1 \AA\ can be powered solely by stellar emission. Theoretically, this is a difficult question to address, since several uncertainties remain in our modelling of pAGB evolution, including the initial-final mass relation and the residual envelope mass \citep{O'Connell1999, Stanghellini2000}.

In this section we compute the time evolution of the EW(H$\rm\alpha)$ after a starburst predicted by population synthesis models (by C. Maraston) including a treatment of the pAGB phase. The stellar population models are the same as those predicting absorption-line indices including abundance-ratio effects, which are required to describe the absorption spectra of old galaxies and bulges \citep{Thomas2004, Thomas2011}. The pAGB phase is included following the stellar tracks for the post-AGB phase based on the calculations of \cite{Stanghellini2000}, and make use of the fuel consumption theorem to compute the post-main sequence evolution, as detailed in \cite{Maraston1998, Maraston2005}. In this context, the \textit{fuel} consumed by a specific phase of stellar evolution is defined to be the amount of stellar mass to be converted in luminosity (where the helium mass is suitably scaled with respect to hydrogen to take into account the different energy release, \citealt{Maraston1998}).

In Fig. \ref{fig3.2} we show the predicted ionising photon flux and EW(H$\rm\alpha)$ for this model. The reference model shows the time evolution of a starburst, demonstrating the precipitous decline in the ionising photon flux and EW(H$\rm\alpha)$ following the death of massive OB stars. The inclusion of the pAGB contribution generates ionising photons in old populations as soon as the pAGB phase contributes a sizeable amount of fuel. We note that the exact time after which the pAGB phase starts consuming sizeable amounts of fuel ($\sim$ 2 Gyr in this model) is strongly dependant on the details of the calculation and the adopted stellar tracks. We plan to explore other evolutionary computations in the future, however, for the results of the present paper, the exact behaviour around 1 Gyr is not relevant. The reference pAGB model produces an ionising photons flux of $\rm Q_{ion} \sim 10^{41} (M_{\star}/M_{\odot})$ and an EW(H$\rm\alpha) \sim 0.8 \AA$. These values are consistent with previous modelling of the pAGB phase presented in the literature \citep{Binette1994, CidFernandes2011}.

In order to gain some physical understanding of the uncertainties involved in the values calculated, we also calculate a simple ad-hoc pAGB prescription, where the ionising flux of the evolved population depends on two free parameter: the fuel consumed during the pAGB phase and the \textit{mean} effective temperature of pAGB stars (where the average is intended to be a time-average; this approach was already followed in \citealt{Maraston2000} to model the UV spectra of elliptical galaxies). Either a small increase in the fuel consumption or an increase in the mean temperature with respect to those assumed in the reference model would increase the predicted  $\rm EW(H\alpha)$. A realistic choice of parameters (fuel in the range $\rm 6\cdot 10^{-3} - 6\cdot 10^{-2} ~ M_\odot$ and mean temperature in the range $\rm 25000 - 40000 ~K$) leads to $\rm EW(H\alpha)$ in the range 0.5-3.0 \AA\ (represented by the shaded region in Fig. \ref{fig3.2}).

 \subsection{LI(N)ERs host old stellar populations}
\label{sec3.3}

\begin{figure*} 
\includegraphics[width=0.8\textwidth, trim=0 160 0 160, clip]{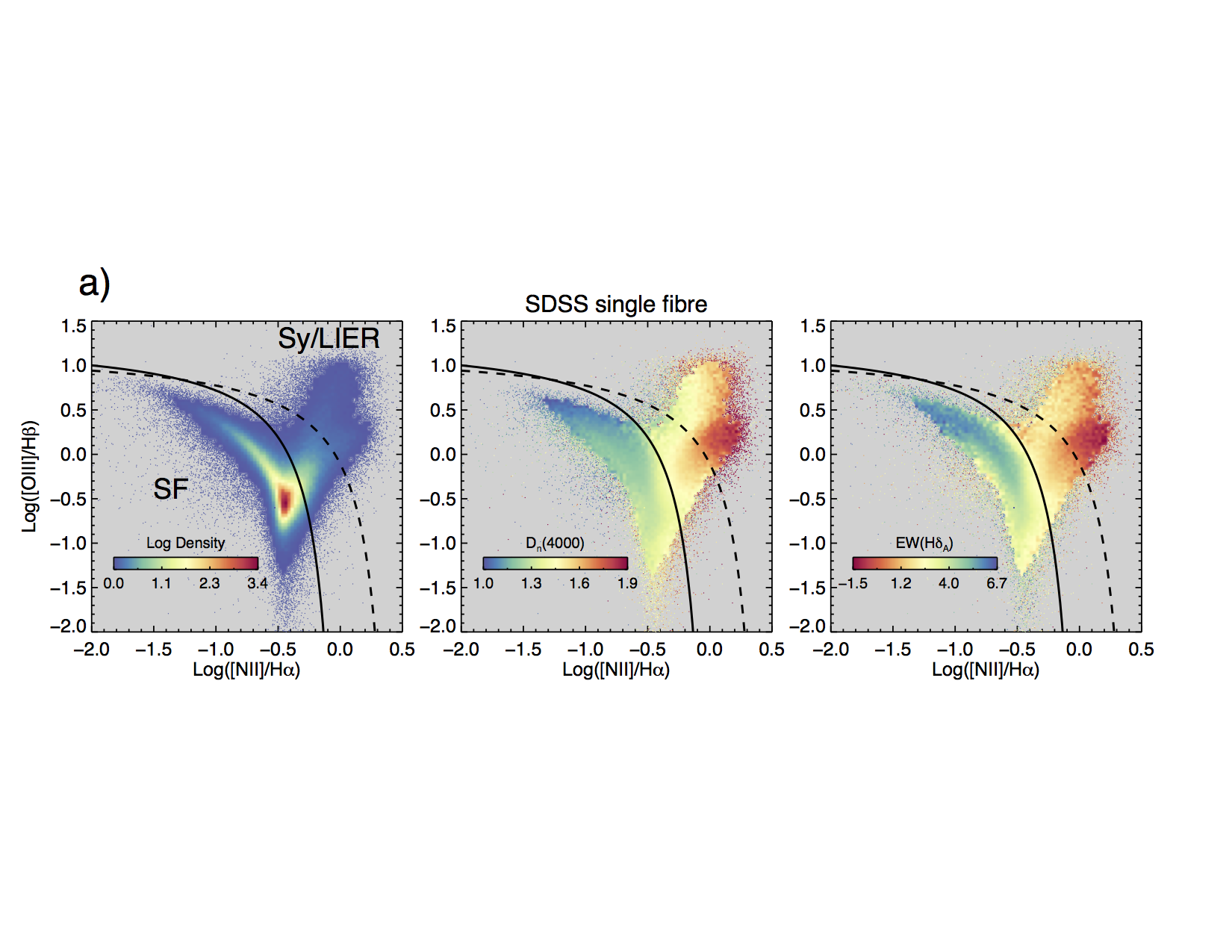}
\includegraphics[width=0.8\textwidth, trim=0 160 0 160, clip]{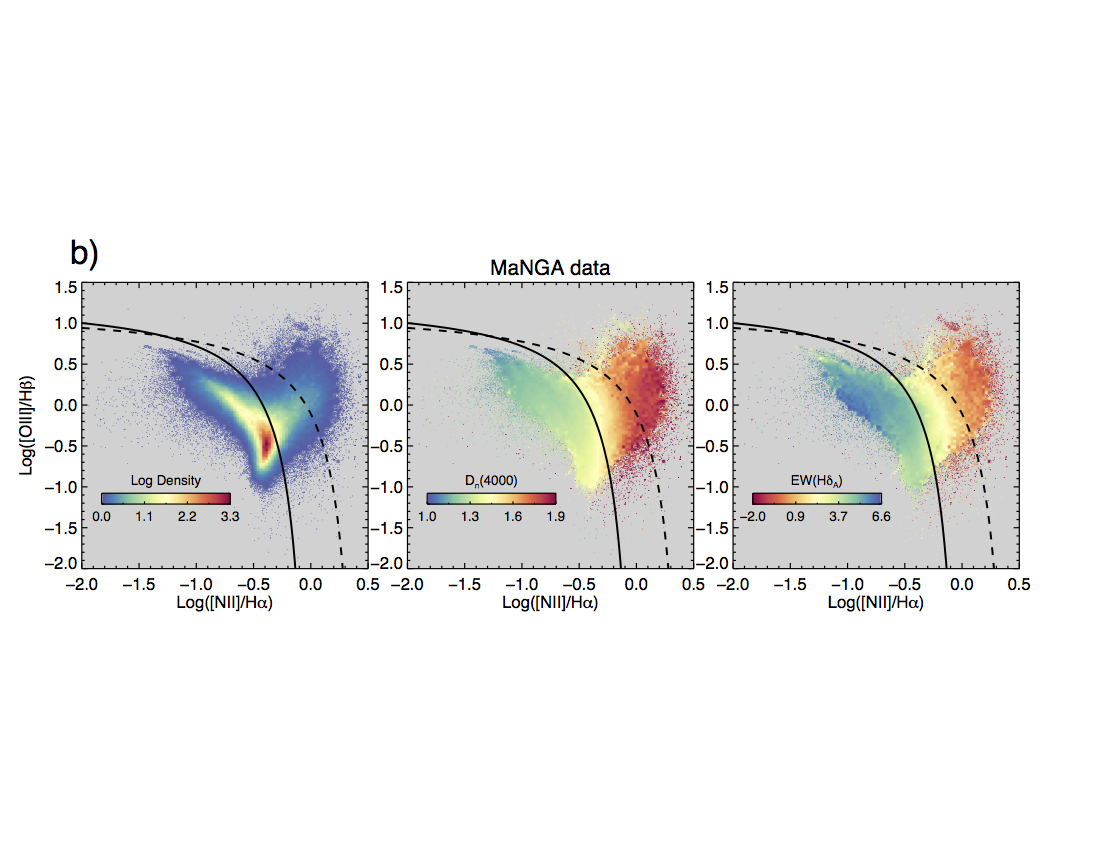}
\includegraphics[width=0.8\textwidth, trim=0 160 0 160, clip]{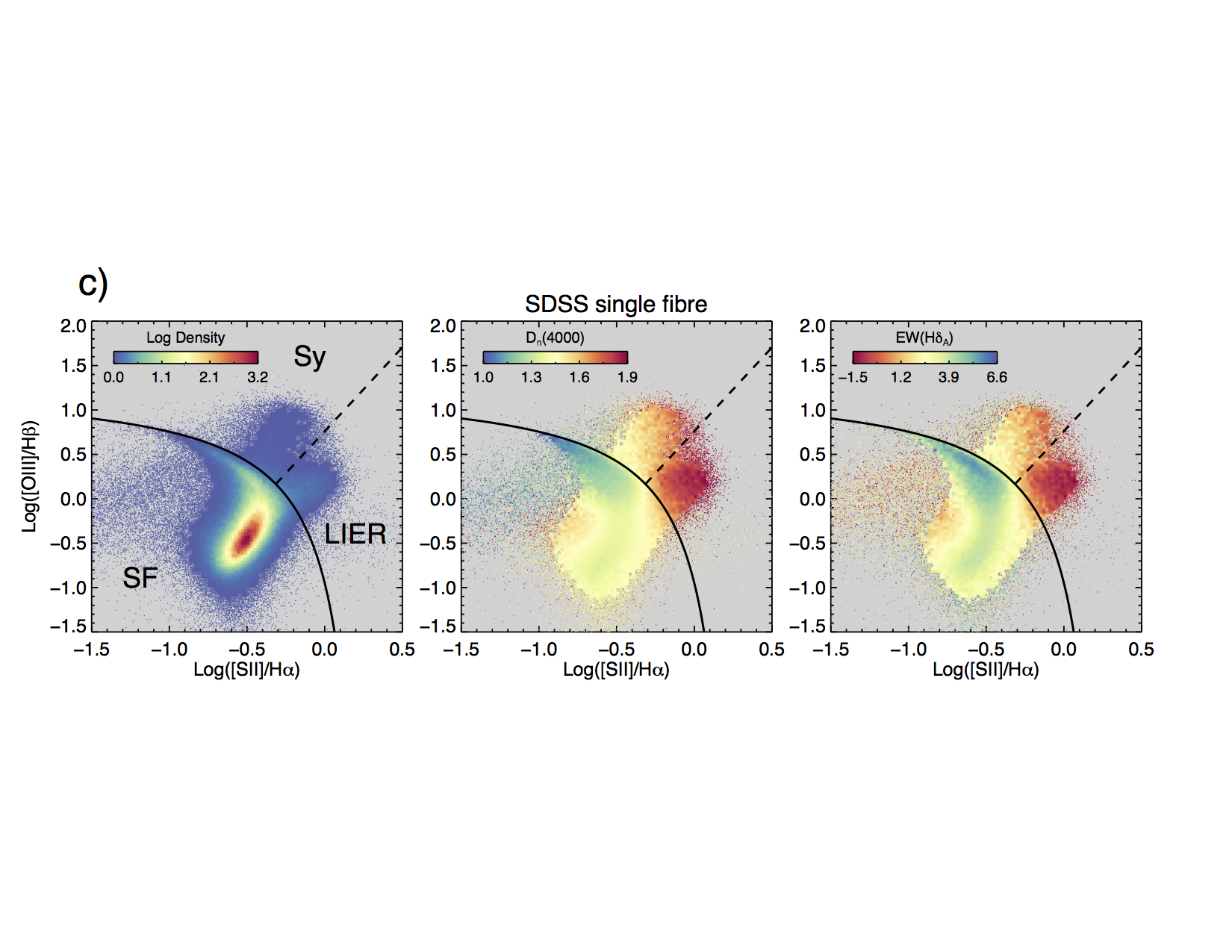}
\includegraphics[width=0.8\textwidth, trim=0 160 0 160, clip]{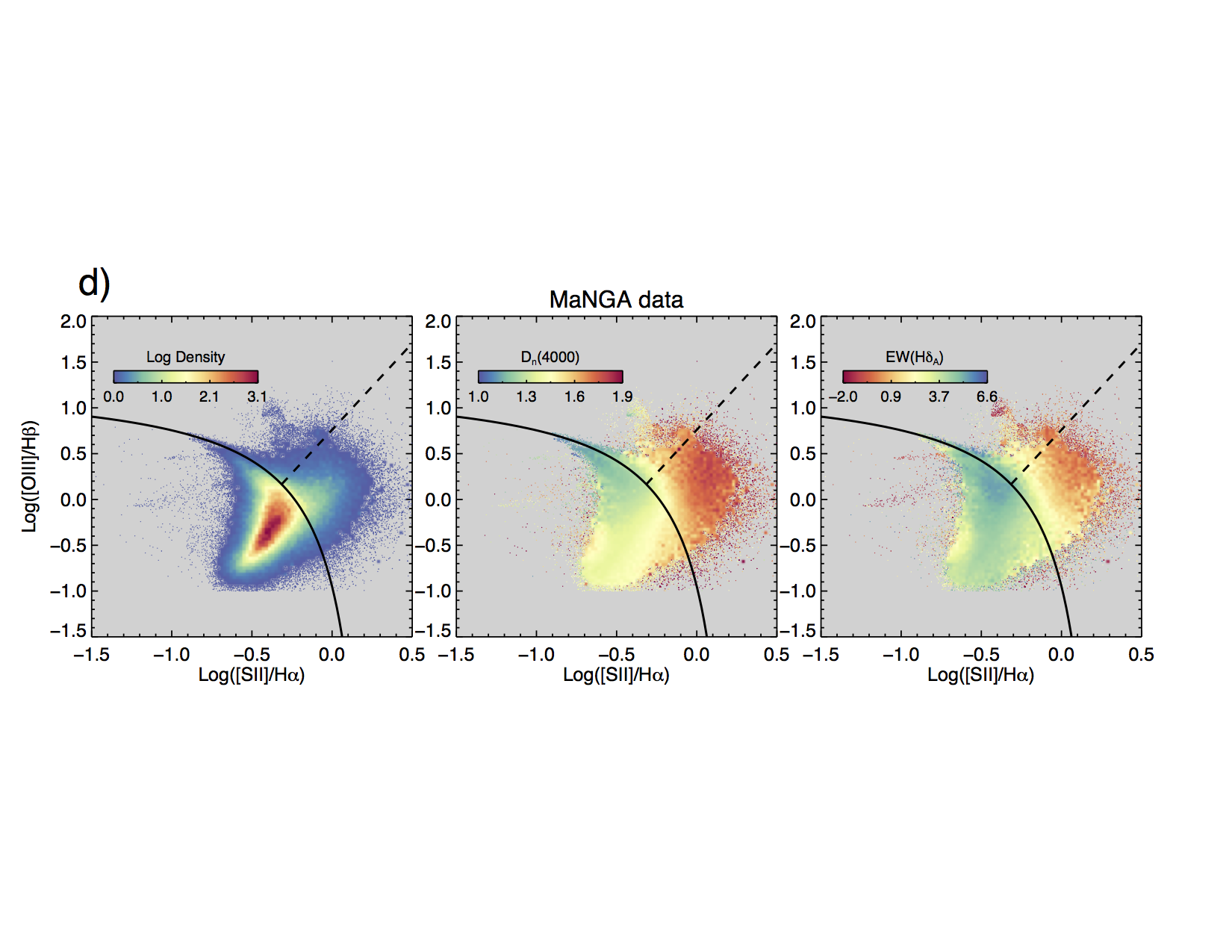}
\caption{The BPT diagnostic diagrams based on [NII]/H$\alpha$ (top) and [SII]/H$\alpha$ (bottom) for the galaxies in the MPA-JHU legacy SDSS sample and MaNGA sample used in this work. The three panels in each row show respectively the density of galaxies (left), the average $\rm D_N(4000)$ (middle) and $\rm EW(H\delta_A)$ (right) across the BPT diagram. The demarcation liens are discussed in Sec. \protect\ref{sec4}. For each plot, both axes have been divided in 100 intervals thus creating small bins in the 2D space. Bins containing more than 20 galaxies are represented by the average value of the quantity (or total number density). For bins containing less than 20 galaxies, the individuals values of each galaxy are shown. For the SDSS legacy data, emission line fluxes and stellar population indices are taken from the MPA-JHU catalogue \protect\citep{Brinchmann2004, Tremonti2004}.} 
\label{fig3.3}
\end{figure*}

In further support of the stellar hypothesis for LI(N)ER emission legacy SDSS spectroscopy has shown that LI(N)ER galaxies have substantially older stellar populations than any other spectral type \citep{Kauffmann2003a, Kewley2006}, including Sy galaxies. In order to allow for a direct comparison between large spectroscopic samples like legacy SDSS and the new MaNGA data, we show in Fig. \ref{fig3.3}a,c the BPT diagnostic diagrams for the SDSS main galaxy sample. Emission line fluxes and spectral indices are taken from the MPA-JHU catalogue \citep{Brinchmann2004, Kauffmann2003a, Tremonti2004}, and the demarcation lines shown are from \cite{Kewley2001} and \cite{Kauffmann2003a}. 

The middle and right panels of Fig. \ref{fig3.3} show the average value of the age-sensitive spectral indices $\rm
D_N(4000)$ and  $\rm H \delta_A$ as a function of position in the BPT diagram. $\mathrm{D_N(4000)}$  (also referred to as
the `$\rm 4000 \ \AA$ break') is calculated here by adopting the definition in \cite{Balogh1999}. The break results from the combination of the Balmer break with a number of metal lines in a narrow wavelength region. In hot stars, these metals are multiply ionised and the break will be small (and dominated by the Balmer break), while as cooler stars start to dominate the stellar population the break increases due to metal-line blanketing. Therefore $\mathrm{D_N(4000)}$ is a sensitive probe of the overall ageing of the stellar population, and is roughly independent of metallicity for ages less than 1 Gyr. The equivalent width of $\mathrm{H \delta_A}$ (defined by \citealt{Worthey1997}) is primarily sensitive to stars of intermediate spectral type (A to early F) and thus increases until about 300-400 Myr after the burst (post-starburst population) and then decreases again at later times \citep{Kauffmann2003b, Delgado2005}. At late ages, both indices are known to depend both on the age and the metallicity of the stellar population \citep{Thomas2004, Korn2005, Thomas2006}.

As evident from Fig.~ \ref{fig3.3}a,c, LI(N)ER emission in the central region of galaxies, as probed by legacy SDSS, is associated with high $\rm D_N(4000)$ values and low EW(H$\delta_A$), hence ruling out the presence of young stars (younger than 1 Gyr) in LI(N)ER regions.

\subsection{The role of diffuse ionised gas (DIG)}
\label{sec3.4}

Diffuse ionised gas (DIG) with LINER-like emission is observed surrounding the plane of the Milky Way and other spiral galaxies \citep{Hoopes2003, Rossa2003, Oey2007, Haffner2009, Blanc2009, Flores-Fajardo2011}.
This gas is warm ($10^4$K), low density ($\mathrm{10^{-1} cm^{-3}}$) and has low ionisation parameter.
Its ionisation source is not yet well understood, but is generally compatible with photoionisation by radiation from stellar sources (O and B stars, but possibly also hot evolved stars), filtered and hardened by photoelectric absorption by gas in the disc. Study of the DIG in external galaxies remains difficult because of its low surface brightness, however extensive work on edge-on galaxies (see for example the prototype case of NGC 891 as studied in \citealt{Rand1990, Rossa2004, Bregman2013}) has demonstrated that DIG emission is detectable up to several kpc from the disc midplane.

Gas with LINER-like emission is also observed in the inter-arm regions of spiral galaxies \citep{Walterbos1994, Oey2007}. Such gas may have an origin similar to the extra-planar DIG, i.e. ionised by radiation escaping H\textsc{ii} regions, filtered and hardened by photoelectric absorption \citep{Mathis2000, Hoopes2003}; however, it may also be associated with in-situ ionisation by the evolved old stars (Zhang et al., \textit{in prep.}).

\subsection{From LINER to LIER}
\label{sec3.5}

Evidently several different physical processes can be responsible for LI(N)ER-like emission in galaxies. 
Studies of the central few parsecs in nearby galaxies \citep{Ho1997, Shields2007} demonstrate that LINER emission on those scales can be associated with nuclear accretion phenomena, in particular weak, radiatively inefficient AGNs (see for example the review from \citealt{Ho2008}). However, LI(N)ER-like emission on larger scales is far from trivial to interpret. With few exceptions (e.g. NGC 1052, \citealt{Pogge2000}) AGN-related LINERs do not show elongated structures analogous to ionisation cones in Sy galaxies.

Moreover, at the typical scales probed by the SDSS fibre and by the MaNGA PSF (few kpc) the predicted H$\alpha$ emission from pAGB stars is comparable to or higher than that expected from a weak AGN. To perform an order of magnitude estimate, let us assume a specific ionisation flux from pAGB stars of $\rm Q_{ion} = 10^{41} (M_{*}/M_{\odot})$ (in line with the models in Sec. \ref{sec3.2}) and an average stellar mass surface density of $\rm \Sigma_{*}=10^3 \ M_{\odot} \ pc^{-2}$ (typical of elliptical galaxies or bulges). Under these assumptions, and assuming that all the ionising photons are absorbed in the gas, a $\rm 1 \ kpc^2$ region will radiate $\rm L_{H\alpha} \sim 10^{38} erg\ s^{-1}$. This is of the same order of magnitude as the typical H$\alpha$ luminosity of well-studied low-accretion rate AGN classified as LINERs (for example, \citealt{Ho2008} quotes an average H$\alpha$ luminosity for LINER AGN of $\rm L_{H\alpha} \sim 5.0 \cdot10^{38} erg\ s^{-1}$). Therefore, on the scales probed by MaNGA, LINER-type AGN photoionisation will only be energetically dominant in the very central regions, if at all. 

This fact has two important consequences: 1) kpc-resolution surveys (like SDSS or MaNGA) are not suitable for detecting and studying low-luminosity (LINER-type) AGN via their optical line emission, 2) the attribute `nuclear' in the LINER acronym is misleading when not referred to true nuclear ($\sim 100$ parsec size) scales. In recognition of the latter, for the rest of this work we drop the `N' in LINER and refer to low ionisation emission-line regions as `LIERs'\footnote{Coincidentally, one might argue this is a more fitting-sounding acronym, in recognition of the long-lasting misclassification of these sources as AGN in surveys like legacy SDSS.}.

\section{MaNGA - excitation morphologies of emission line galaxies}
\label{sec4}

\begin{figure*} 
\includegraphics[width=1.0\textwidth, trim=0 150 0 150, clip]{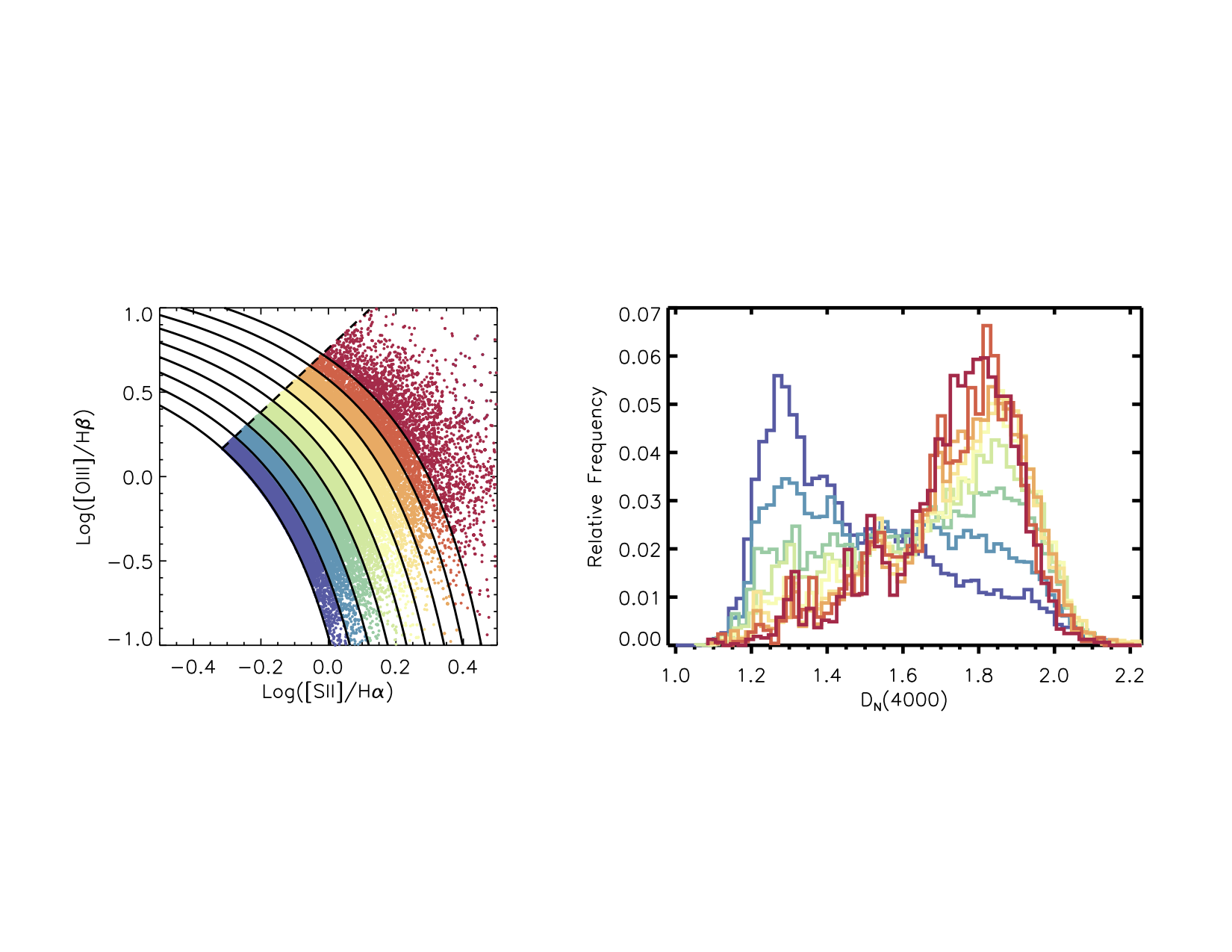}
\caption{A zoom-in in the LIER region of the [SII] BPT diagram. Following \protect\cite{Kewley2006}, we draw a set of curves parallel to the demarcation curve of \protect\cite{Kewley2001} to define the distance from the SF sequence. Successive curves are separated by a 0.05 dex interval in $\rm log([SII]/H\alpha)$ and $\rm log([OIII]/H\beta)$. The right panel shows the histograms of the $\rm D_N (4000)$ distribution as a function of distance from the SF sequence (the colour-coding is the same as in the left panel, with red being furthest from the star forming sequence). Within the LIER spaxels, a clear transition is observed in $\rm D_N (4000)$ as a function of distance from the SF sequence.}
\label{fig4.1}
\end{figure*}

\subsection{The BPT diagram on kpc scales}
\label{sec4.1}

In \cite{Belfiore2015a} we have used preliminary MaNGA data (P-MaNGA) to study the resolved properties of the ionised gas for a small sample of 14 galaxies observed with a prototype of the MaNGA instrument. In this work we present a more detailed analysis of the resolved excitation properties of the current, much larger, MaNGA galaxy sample. In Fig. \ref{fig3.3}b,d we show the BPT diagnostic diagrams for all the individual spaxels in the current MaNGA sample which have S/N $>$ 2 in all the relevant strong emission lines and S/N $>$ 5 on the continuum. We have checked that similar results are obtained using a different S/N cut or a cut on the error of the considered line ratios \citep{Juneau2014}.\footnote{This cut selects approximately $3 \cdot 10^5$ spaxels. Note, however, that spaxels in MaNGA are not statistically independent due to the effects of the PSF and the datacube generation algorithm. On average, considering nearby spaxels, there are approximately 20 spaxels per each PSF element.}

The left panels show the density distribution of all the MaNGA spaxels on the [NII] and [SII] BPT diagrams. In both diagrams the star forming (SF) sequence is clearly visible, together with the less populated right-hand branch. As will be discussed in detail in the following sections (and demonstrated in \citealt{CidFernandes2011}), the S/N cuts imposed above have a strong effect on the fraction of LIER-like spaxels which can be plotted in this diagram, since LIER-like emission is characterised by low surface brightness of line emission. 

In the [NII] BPT diagram, the solid line represents the empirical demarcation line suggested by \cite{Kauffmann2003a} to encompass the sequence of star forming galaxies. The dotted line is based on the photoionisation models of \cite{Kewley2001} and represents the upper envelope of their H\textsc{ii} regions models, implying that any point above the line needs to have additional contributions from other ionisation sources. However, it should be noted that more recent models have found that star forming galaxies can be characterised by line ratios extending beyond such demarcation \citep{Feltre2015}. Also, one should take into account that such a diagram is affected by variations of the N/O abundance ratio, which can vary significantly from galaxy to galaxy depending on the star formation history. The [SII] BPT diagram is not affected by the N/O abundance ratio, and provides a cleaner separation within the `right-hand sequence' between Sy and LIERs \citep{Kewley2006, Schawinski2010}. In this work we adopt the demarcation line between the SF and the right-hand sequence from \cite{Kewley2001}, while the Sy-LI(N)ER (dotted in figure) demarcation line is the empirical demarcation line presented in \cite{Kewley2006}.

The middle and right panels of Fig. \ref{fig3.3}b,d show the same sample of spaxels colour-coded by the mean $\rm D_N (4000)$ and EW$\rm (H\delta_A)$ in each bin. The `right-hand' sequence in the [NII] BPT diagram and the LIER space in the [SII] BPT diagram show stellar indices typical of old stellar populations (high $\rm D_N (4000)$ and low EW$\rm (H\delta_A)$), similarly to what is observed in legacy SDSS (Fig. \ref{fig3.3}a,c). Given the adopted S/N threshold, using Monte Carlo simulations we estimate the maximum uncertainties for spectral indices to be respectively 0.03 for $\rm D_N (4000)$ and 0.5 \AA\ for EW$\rm (H\delta_A)$.

Compared to the single fibre SDSS data, in the spatially resolved MaNGA data young stellar populations (i.e. low $\rm D_N(4000)$ and high EW(H$\rm\delta_A$)) spread towards larger values of [SII]/H$\alpha$, [NII]/H$\alpha$, and larger [OIII]/H$\beta$. The result is that young stellar populations extend slightly beyond the adopted demarcation lines between star forming galaxies and AGN/LIERs, in particular in the [SII] BPT diagram, suggesting a potential misclassification of SF regions as dominated by AGN photoionisation in this area of the BPT diagram. Intriguingly, this effect has also been recently observed in high-spatial resolution IFS observations of a low-metallicity dwarf galaxy with VLT-MUSE \citep{Fensch2016}.

In Fig. \ref{fig4.1} we focus on the LIER area of the [SII] BPT diagram and draw a set of parallel lines (distanced by 0.05 dex) to define an effective distance from the star forming sequence (compare with a similar study of SDSS LIERs in \citealt{Kewley2006}). The right panel of Fig. \ref{fig4.1} demonstrates the existence of two populations of LIER regions with distinct stellar populations. At small distances from the SF sequence, LIER emission is characterised by young ages ($\rm D_N (4000) \sim 1.3$), while moving outwards from the SF sequence we see the emergence of a second populations of LIER regions with much older stellar populations ($\rm D_N (4000) \sim1.9$). The younger age LIER population is virtually absent in single fibre SDSS data, and is therefore only seen with the larger radial coverage and spatial resolution of MaNGA. We will see in the next sections that this population of LIER regions is mainly associated with diffuse ionised gas and extra-planar emission in star forming galaxies. 

We note in passing that the Sy branch is underpopulated in MaNGA with respect to legacy SDSS. This is due to a combination of selection effects. Importantly, legacy SDSS observes at higher mean redshift and has much superior statistics in terms of sampling galactic central regions, where AGN ionisation is likely to dominate, while MaNGA employs a large number of its fibres to target galactic outskirts.

\subsection{A new classification scheme based on spatially resolved line emission}
\label{sec5.1}

Simple demarcation lines in the BPT diagrams have been the main classification tool to distinguish star formation from other sources of ionisation in several studies based on the SDSS and more distant galaxy samples \citep{Kauffmann2003a, Tremonti2004, Kewley2006, Kewley2013, Shapley2015}. However, a large amount of previous investigations using IFS (see for example \citealt{Rich2010, Sharp2010, Singh2013, Davies2014, Belfiore2015a, Gomes2015b}) have paved the way to the realisation that emission line classifications based on single fibre spectroscopy can be systematically biased. In particular, previous IFS work has highlighted the diversity of the ionisation structure within single galaxies. Such diversity is particular dramatic in `peculiar' galaxies, such as mergers and galaxies hosting AGNs. But `regular' galaxies can also be characterised by different excitation processes in different galactic regions. In this section we propose a systematic classification of the `excitation morphologies' of galaxies, based on the large MaNGA sample currently available. The philosophy behind this scheme is to represent a natural extension of the classifications based on single-fibre spectroscopy, while attempting to correct for significant aperture biases.

\begin{figure*} 
\includegraphics[width=\textwidth, trim=10 0 0 0, clip]{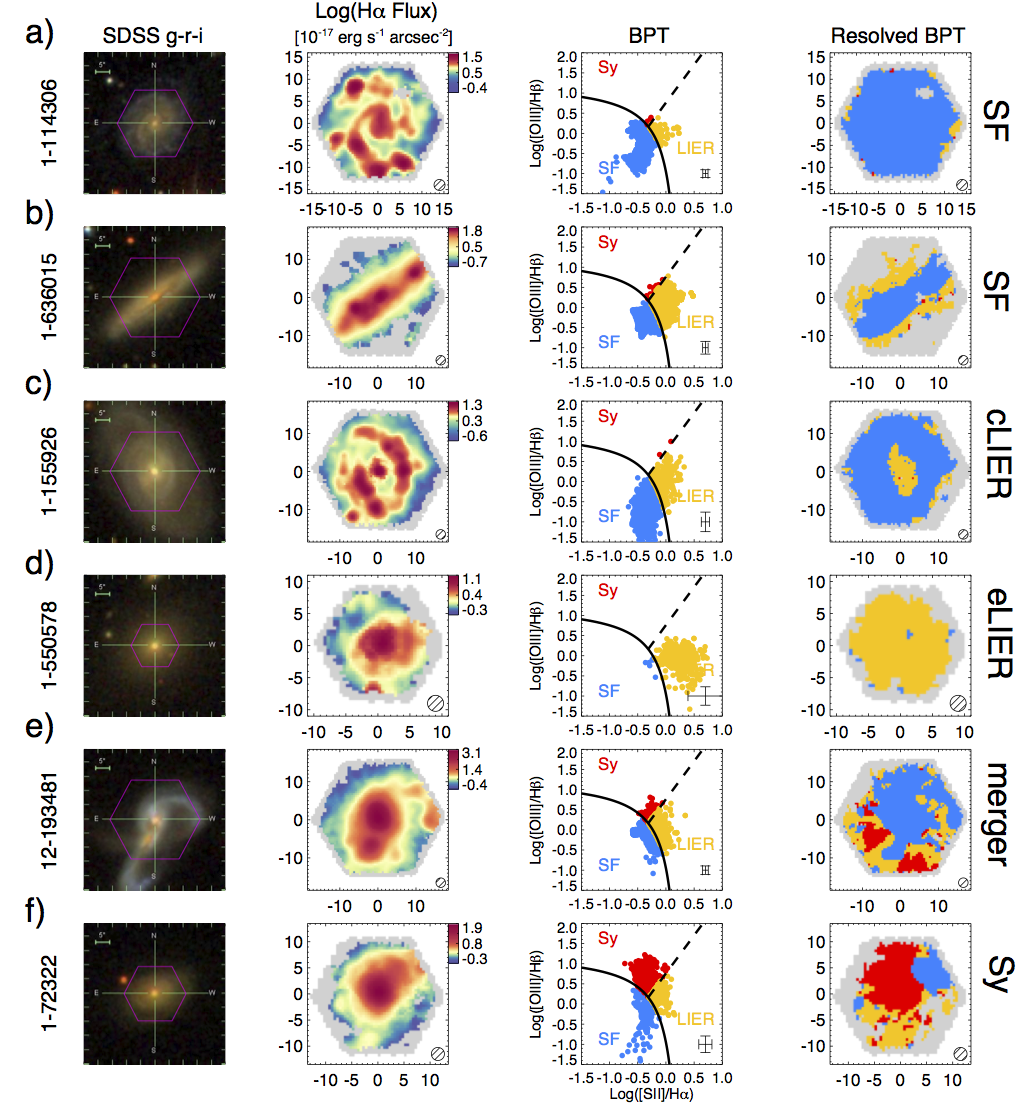}
\caption{Examples of the different gas excitation conditions observed in MaNGA galaxies. Following the classification scheme described in the text, from top to bottom, galaxies are classified as follows: {\tt1-114306} star forming (SF); {\tt1-636015} star forming (note the clear extra-planar LIER emission); {\tt1-155926} LIER central (cLIER); {\tt 1-114928} LIER extended (eLIER); {\tt 12-193481} peculiar (merger); {\tt 1-72322} Seyfert (Sy). Each row shows, from right to left 1) a $g-r-i$ image composite from SDSS with the MaNGA hexagonal FoV overlaid 2) A map of Log$(\rm H \alpha$ Flux$)$, with histogram stretching of the colour-bar. 3) the [SII] BPT diagram showing the position of each spaxel (and typical line ratio uncertainties in the bottom right corner). 4) A map of the galaxy colour-coded according to the position of individual regions with the same colour-coding as in 3). The MaNGA PSF is shown as a hatched circle in the bottom right corner of the MaNGA maps. The greyed area corresponds to the FoV of the MaNGA bundle.}
\label{fig5}
\end{figure*}

In Fig. \ref{fig5} we illustrate with some prototypical examples the types of ionisation structures observed in MaNGA galaxies and our proposed revised classification scheme.

\subsubsection{Star forming (SF) galaxies}

The galaxy in Fig. \ref{fig5}a (MaNGA-ID {\tt 1-114306}) is a typical star-forming galaxy. Its prominent spiral arms are populated by bright star-forming regions, which at the MaNGA kpc resolution generally consist of several classical H\textsc{ii} regions \citep{Mast2014}, evident in the H$\alpha$ flux map. The vast majority of spaxels are classified as star forming using either the [SII] and [NII] BPT diagrams. 

Galaxy {\tt 1-636015} (Fig. \ref{fig5}b) shows the typical ionisation structure of strongly star forming edge-on discs. While the disc itself is characterised by SF line ratios, moving above and below the disc we observe a transition to LIER excitation. The change in excitation conditions can be interpreted as a transition from H\textsc{ii}-dominated ionisation in the disc to DIG, as discussed in Sec. \ref{sec3.4}, and possibly shocks \citep{Fogarty2012, Ho2016}. We note that LIER emission consistent with DIG is also sometimes observed in MaNGA in SF galaxies in inter-arm regions. Further work in this series will be dedicated to the detailed study of the ionisation properties as a function of inclination and surface brightness of the DIG.

For the purpose of this work, since we aim at a general galaxy classification scheme, LIER-like DIG is considered a constituent part of the ionised gas in normal SF galaxies. We therefore do not assign a different class to SF galaxies where LIER-like DIG is detected in the outer/inter-arm regions. Thus, operationally a galaxy is defined as SF if a region of the size of the MaNGA PSF in the galaxy centre is classified as star forming.

\subsubsection{Central LIER (cLIER) galaxies}

The galaxy in Fig. \ref{fig5}c (MaNGA-ID {\tt 1-155926}) exemplifies a galaxy class showing LIER emission at small galactocentric radii and a transition to SF line ratios further out. Observations of the nuclear regions (e.g. with SDSS single fibre spectroscopy) would classify this galaxy as LI(N)ER, but would miss the star formation in the disc. We denote this class of galaxies `central LIER' (cLIER). In order to define a galaxy as cLIER, we require the LIER emission in the centre to be spatially extended (i.e. larger than the MaNGA PSF, indicated in the bottom-right corner of the maps in Fig. \ref{fig5}), and at the same time star forming regions to be clearly detected at larger radii (larger than one PSF). After an initial automatic classification the maps for all galaxies were visually inspected to confirm the robustness of the classification to the effect of noise/marginal line detections. As a class of galaxies cLIERs have not yet been subject of a systematic investigation in the literature, although they can be related to the small samples presented in previous work \citep[e.g.][]{Gomes2015a, Gomes2015b, James2015p}.

\subsubsection{Extended LIER (eLIER) galaxies}

The galaxy in Fig. \ref{fig5}d (MaNGA-ID {\tt 1-550578}) is a typical example of an early type galaxy showing extended LIER emission. This is the class of LI(N)ERs that has received the most attention in the literature, especially through the contributions of the \textsc{SAURON} and \textsc{ATLAS}$\rm^{3D}$ projects \citep{Sarzi2006, Sarzi2010} and more recently the CALIFA team \citep{Kehrig2012, Singh2013, Papaderos2013}, and for which the non-AGN nature of LIER emission has the strongest support. We will refer to these galaxies as `extended LIER' (eLIER). We consider a galaxy to be an eLIER if it does not show signs of star formation at any galactocentric distance and its central regions are classified as LIER.

\subsubsection{Mergers and interacting systems}

In order to highlight the complications arising in `peculiar' systems, we show the MaNGA observations of a major merger of two star forming galaxies in Fig. \ref{fig5}e (MaNGA-ID {\tt 12-193481}). The central regions and most of the tails of the individual merging galaxies are dominated by SF line ratios, while a large fraction of the more diffuse gas shows line ratios falling beyond the SF sequence in the BPT diagram (both LIER-like and Sy-like). Shocks, non-virial motions and outflows are likely to play a major role in the ionisation structure of this type of systems (see further discussion in Sec. \ref{sec8.3}). For the purpose of this work we exclude mergers from our sample of `normal' galaxies.

\subsubsection{Seyfert galaxies}

Seyfert galaxies (Sy) generally display large regions photoionised by the nuclear source (photoionisation cones), extending from the nucleus to kpc scales, often associated with disturbed gas kinematics (outflows and other non-virial motions). In Fig. \ref{fig5}f (MaNGA-ID {\tt 1-72322}) we show an example of a Type 2 Seyfert galaxy. 
The excitation structure appears complex, featuring LIER ratios and also SF regions outside the area photoionised by nuclear emission. Seyfert galaxies (i.e. galaxies where the central PSF is classified as Seyfert using the [SII] BPT diagram) are excluded from the analysis presented in the remaining sections of this paper.

\subsubsection{Line-less galaxies}

Finally, some galaxies do not show any detectable significant line emission. We will refer to these galaxies as `line-less'. In the literature, galaxies are often defined as `line-less' if their emission lines do not meet a set S/N level \citep[e.g][]{Brinchmann2004}. However, we consider that a definition which is independent from the depth of the data would be preferable. In this work, therefore, a galaxy is defined as line-less if the mean EW(H$\alpha$) within 1$\rm R_e$ is less than 1 \AA. The chosen EW has the advantage of corresponding closely to the sensitivity limit of the MaNGA data. However, we stress that the chosen numerical value does not have a physical motivation and is chosen purely for convenience. We further comment on the definition of line-less galaxies in Appendix \ref{appA}.

\subsection{Summary of galaxy classes}
\label{sec5.2}

\begin{figure} 
\includegraphics[width=0.49\textwidth, trim=50 30 80 80, clip]{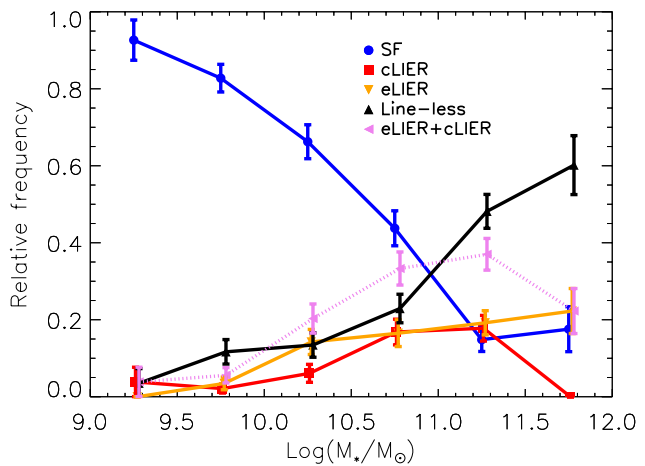}
\caption{The fraction of normal galaxies (excluding interacting galaxies, mergers and AGN) in the current MaNGA sample classified as star-forming (blue), central LIER (red), extended LIER (orange) and quiescent (black) as a function of stellar mass. The sum of both eLIER and cLIER classes is also shown (dotted magenta line). Galaxies with ambiguous BPT classification have also been excluded. The MaNGA sample is selected to be representative of the total galaxy population in each stellar mass bin for $\rm log(M_\star/M_\odot) >9$ and appropriate volume corrections have been applied to the current sample to correct for effect of the MaNGA selection function. Errors shown are the formal counting errors.}
\label{fig5.2}
\end{figure}

The excitation properties of local galaxies, as probed by MaNGA, follow surprisingly regular radial patterns on kpc scales. Apart from the complex ionisation conditions observed in peculiar and active systems (interacting galaxies and Sy), the presence of spatially resolved LIER ionisation increases gradually in importance from SF galaxies to cLIER and eLIER galaxies. Moreover, not all possible combinations of excitation properties are observed. For example, there are no galaxies with a line-less central region and star formation at larger radii, nor face-on\footnote{As described in Sec. \ref{sec5.1}, LIER excitation often dominates at large distances from the disc midplane in edge-on SF galaxies.} galaxies with a SF nucleus and extended areas of LIER excitation in the disc. 

In Fig. \ref{fig5.2} we show the fraction of galaxies in the SF, eLIER, cLIER and line-less classes as a function of stellar mass. Mergers, interacting galaxies, Sy and galaxies with an ambiguous BPT class have been excluded from this plot. The total number of galaxies in each class in the current sample is as follows: SF (314), cLIER (57), eLIER (61), line-less (151), others (including Sy, mergers and unclassified galaxies: 63). The MaNGA sample is not volume-limited as a whole, but it is volume limited in each stellar mass bin with $\rm log(M_\star/M_\odot) > 9.0$. Hence the fraction of galaxies in each class per stellar mass bin are representative of the total galaxy population at $z \sim 0$ (neglecting possible evolution within the MaNGA redshift range $\rm 0.01 < z< 0.15$).

Fig. \ref{fig5.2} reproduces the well-know increase in the red fraction (here generally represented by
quiescent and eLIER galaxies) as a function of mass \citep[e.g][]{Peng2010}. Intriguingly the fraction of
both eLIER and cLIER galaxies is consistent with zero for $\rm log(M_\star/M_\odot) < 10$ and increases
to 15\% in the mass bins $\rm 10.5<log(M_\star/M_\odot) < 11.5$. Overall each of the two LIER galaxy
classes (eLIER and cLIER) represents roughly half of the total number of galaxies that would have been
classified as LI(N)ER using only nuclear spectra. The total LIER galaxy populations (eLIER and cLIER) is
shown as a dotted line and constitutes around $\sim 30 \%$ of the galaxy population for $\rm 10.5 < log(M_\star/M_\odot) < 11.5$, roughly consistent with previous estimates based on single-fibre spectra \citep{CidFernandes2011}. In \textit{Paper II} we present a detailed discussion of the morphology, colours and dynamical properties of each galaxy class. Further examples of galaxies in both LIER galaxy classes are shown in Fig. \ref{fig_ex1} (eLIERs) and Fig. \ref{fig_ex2} (cLIERs) in Appendix \ref{appb}.


\begin{figure*} 
\includegraphics[width=0.49\textwidth, trim=80 80 80 130, clip]{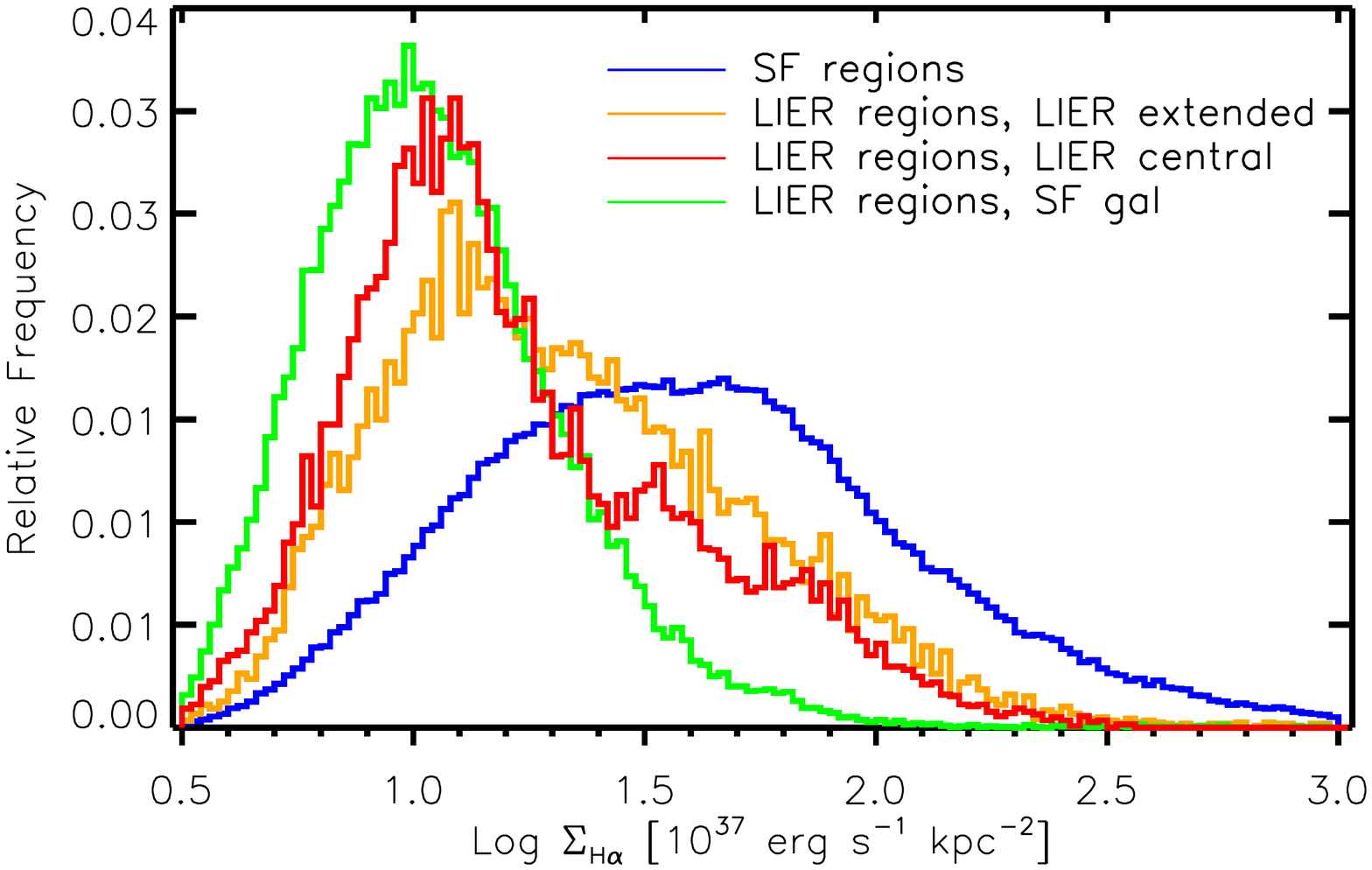}
\includegraphics[width=0.49\textwidth, trim=80 80 80 130, clip]{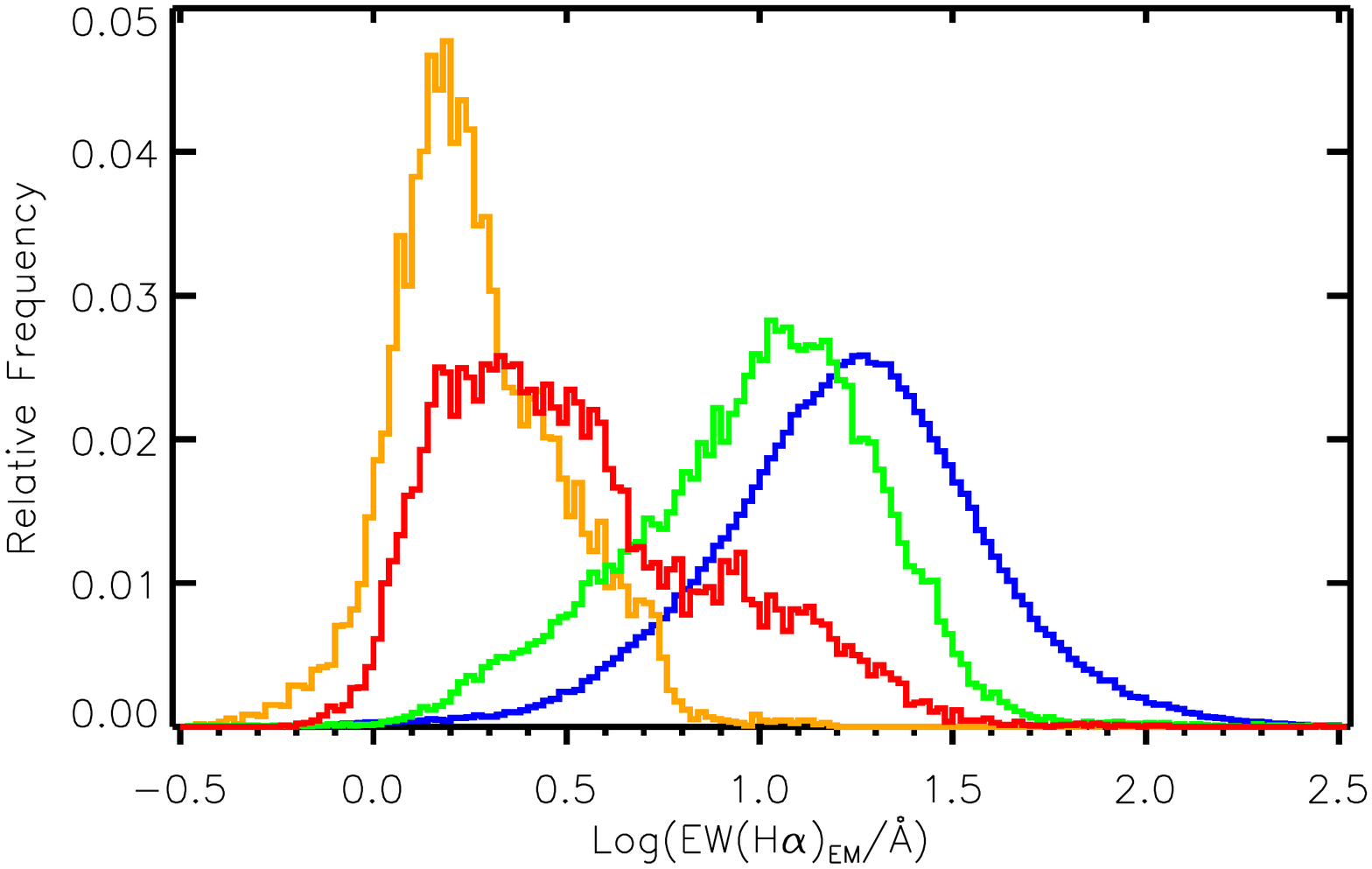}
\caption{Histograms of the distribution of H$\alpha$ surface brightness and the equivalent width of H$\alpha$ (in emission) for MaNGA spaxels. Spaxels of different classes are colour-coded differently following the legend in the left panel. The blue histograms correspond  to spaxels classified as SF (independently of galaxy class). LIER spaxels are subdivided in three classes, depending on the galaxy type: LIER extended (or eLIER, in orange), LIER central (cLIER, in red) and SF galaxies (in green).}
\label{fig6.1}
\end{figure*}

\begin{figure*} 
\includegraphics[width=0.96\textwidth, trim=0 160 0 180, clip]{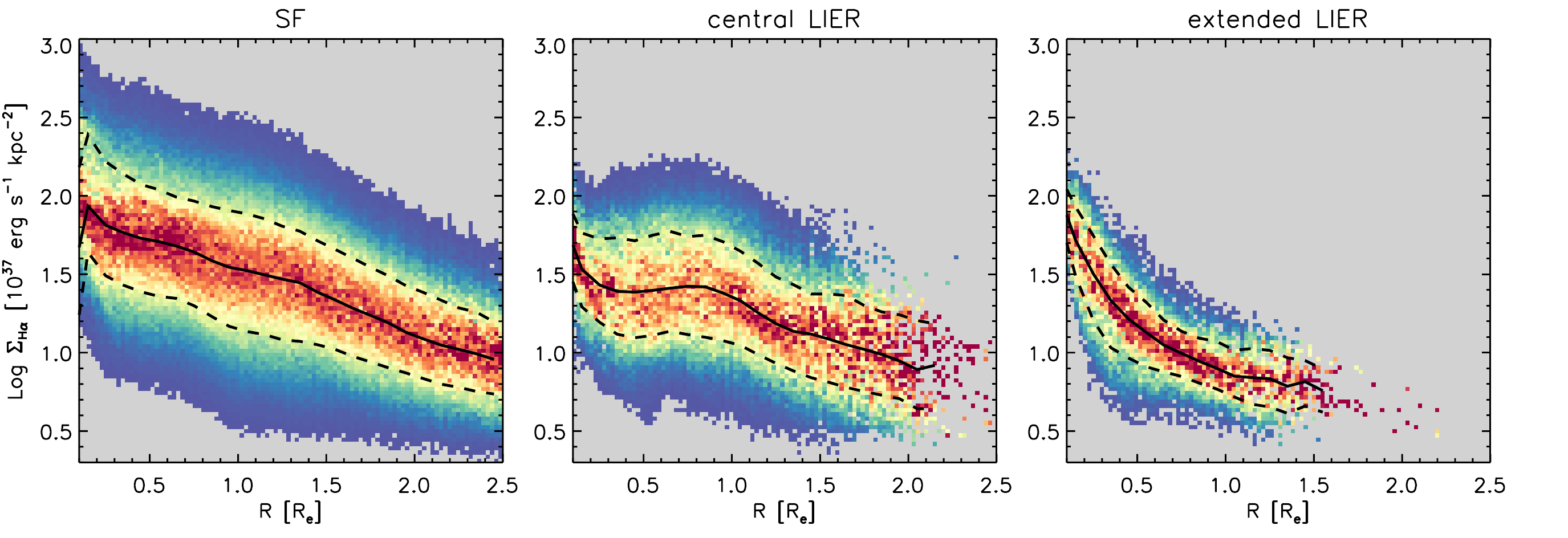}
\includegraphics[width=0.96\textwidth,  trim=0 160 0 195, clip]{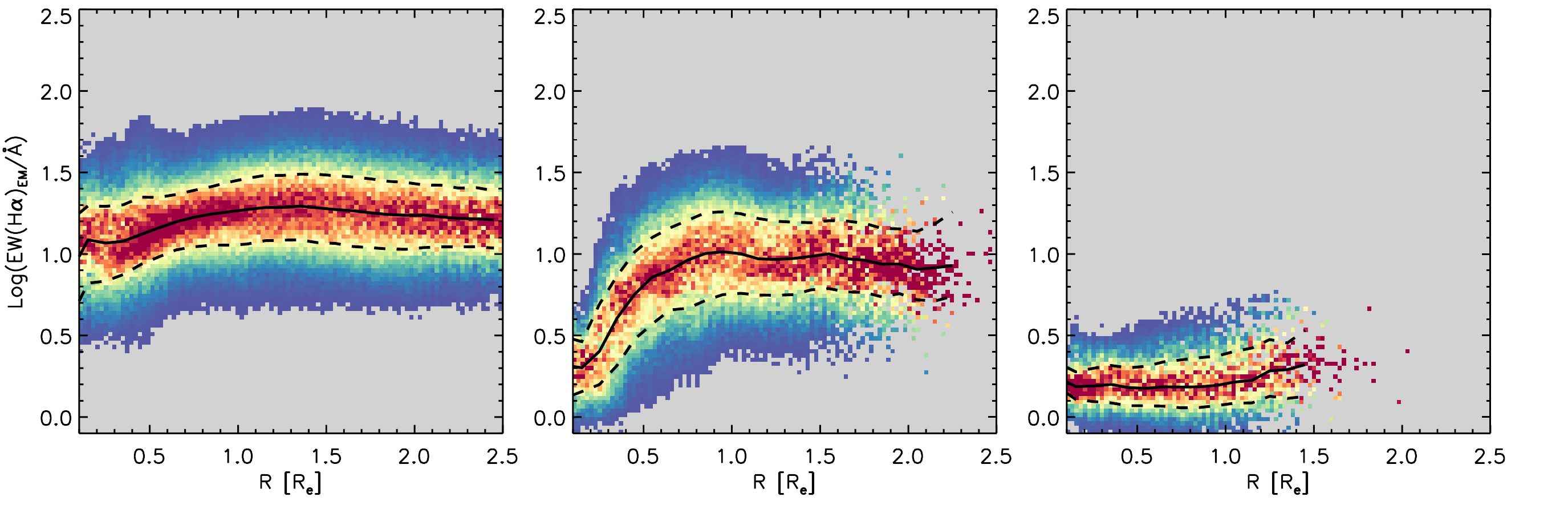}

\caption{Gradients of H$\alpha$ surface brightness (top) and EW($\rm H\alpha$) in emission (bottom) as a function of deprojected radius in units of effective radius. The spaxels are binned in the 2D space and only bins with more than 4 spaxels are shown. The density distributions are normalised at each radius. The black solid line represent the median at each radius, and the dashed lines represent the 25th and 75th percentiles.}
\label{fig6.2}
\end{figure*}

\section{H$\alpha$ emission profiles}
\label{sec6}

\subsection{H$\alpha$ surface brightness and equivalent width distributions}
\label{sec6.1}

The intensity of the Balmer recombination lines of hydrogen is proportional to the ionising photon flux, assuming that all ionising photons are absorbed by the gaseous component. The escape fraction for Lyman continuum photons in late type galaxies in the local Universe is generally very low. Yet there is some evidence that the escape fraction might be larger in early type galaxies, if ionising photons are produced by old stars and the gas distribution is patchy \citep[e.g.][]{Papaderos2013}. Bearing this in mind, we first investigate the distribution of H$\alpha$ surface brightness and equivalent width (EW) as a function of the \textit{local} ionisation conditions. More specifically we divide the MaNGA spaxels in four classes

\begin{enumerate}
\item{SF spaxels in any galaxy type (blue in Fig. \ref{fig6.1}).}
\item{LIER spaxels in SF galaxies (green in Fig. \ref{fig6.1}).}
\item{LIER spaxels in eLIER galaxies (orange in Fig. \ref{fig6.1}).}
\item{LIER spaxels in cLIER galaxies (red in Fig. \ref{fig6.1}).}
\end{enumerate}

This choice is motivated by the observation that H\textsc{ii} regions have similar nebular properties in all galaxies, while the excitation mechanism leading to LIER emission might be different as a function of galaxy class. Both the red and orange histograms refer to regions of spatially resolved LIER-like emission (in cLIER and eLIER galaxies respectively), while the green histogram refers to (mostly extra-planar) diffuse LIER-like emission in SF galaxies. Spaxels that are classified as having Seyfert-like ionisation are not considered in this section.

\begin{figure*} 
\includegraphics[width=0.48\textwidth, trim=20 45 45 45, clip]{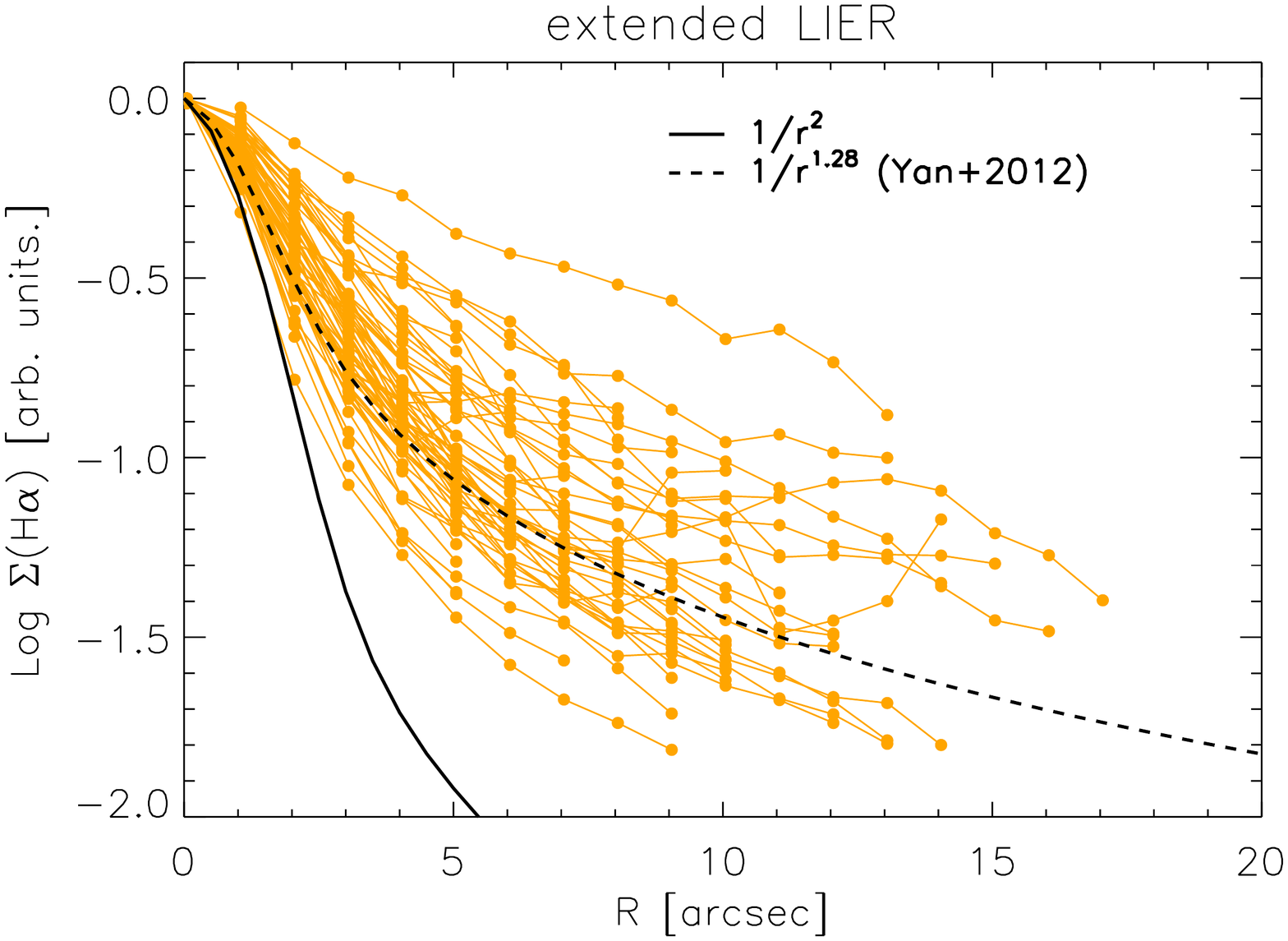}
\includegraphics[width=0.48\textwidth, trim=20 45 45 45, clip]{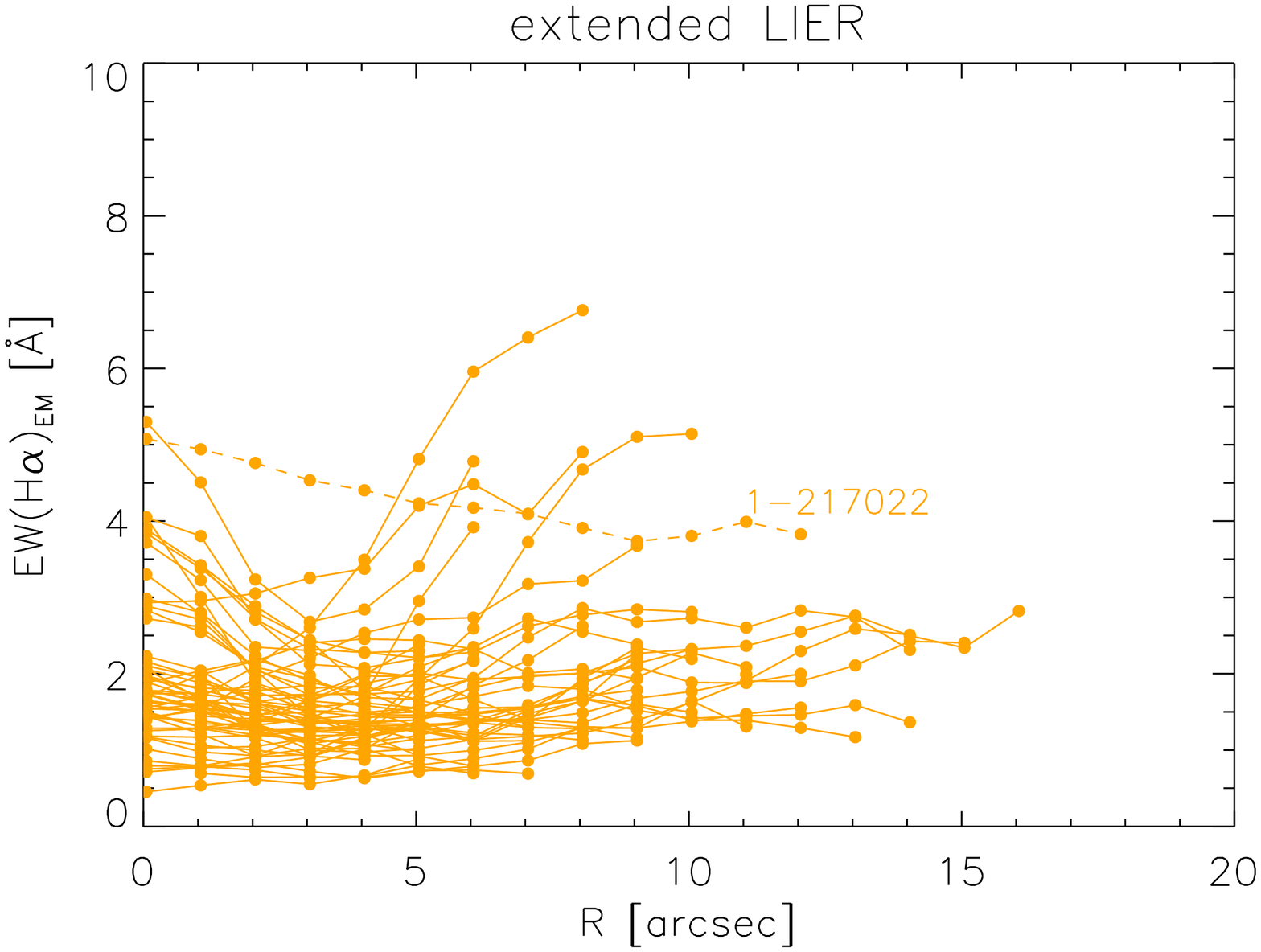}
\caption{H$\alpha$ surface brightness normalised to the central H$\alpha$ flux (left) and EW(H$\alpha$) (right) profile as a function of deprojected radius (in arcsec) for eLIER galaxies. In the left panel, the solid line represents a $\rm 1/r^2$ model convolved with the median MaNGA PSF, representing the prediction from ionisation from a central source assuming constant cloud number density, effective area and filling factor as a function of radius. The dashed line represent the best-fit profile for red LIER galaxies from \protect\cite{Yan2012}, corresponding to $\rm 1/r^{1.28}$. The right panel shows the remarkably flat EW(H$\alpha$) distribution for eLIER galaxies. Galaxy {\tt 1-217022}, standing out as having high EW(H$\alpha$) with respect to the rest of the class, is studied in detail in Cheung et al., \textit{submitted}.}
\label{fig6.3}
\end{figure*}

The distribution of H$\alpha$ surface brightness (left panel of Fig. \ref{fig6.1}) highlights two important facts: 
\begin{enumerate}
\item{SF regions have surface brightness on average 0.4 dex higher than LIER regions, independently of galaxy class.}
\item{LIER regions in any galaxy type have a similar H$\alpha$ surface brightness distribution, with LIER regions in SF galaxies having a slightly lower average surface brightness and less prominent high surface brightness tail. The sharp low surface brightness cutoff for all spaxel classes is due to the sensitivity limit of the MaNGA observations.}
\end{enumerate}

These observations motivate our earlier remark that sensitivity limitations affect more strongly the LIER region of the BPT diagram than the SF sequence. Moreover, even though LIER emission is fainter than line emission due to star formation, if we assume it is present as a diffuse background, it can contribute a non-negligible fraction of the line flux observed in SF regions. We will further quantify the effect of LIER-like DIG in SF galaxies in future work.

The EW(H$\alpha$) has been used by several authors to distinguish between ionisation due to star formation and AGN (characterised by high EW(H$\alpha$)) and ionisation due to a smooth background of hot evolved stars (EW(H$\alpha$)$<$ 3 \AA). Moreover, \cite{CidFernandes2011} have shown that the emission line galaxy population in SDSS is clearly bimodal in EW(H$\alpha$), and that 3 \AA\ is a useful empirical separation between the two populations.

In Fig. \ref{fig6.1} (right panel) we show the EW(H$\alpha$) of the MaNGA spaxels following the classification presented at the start of this section. The distribution of EW(H$\alpha$) for MaNGA spaxels is clearly bimodal, with two independent peaks at log(EW(H$\alpha$)) $\sim$ 0.3 and log(EW(H$\alpha$)) $\sim$ 1.3, in excellent agreement with results from SDSS single-fibre spectroscopy, and clearly separating regions characterised by LIER emission from regions characterised by SF line ratios. We note that the MaNGA sensitivity limit and the arbitrary cut at 1.0 \AA\ used to define line-less galaxies will affect the distribution of low EW spaxels. In Appendix \ref{appA} we discuss a possible way of folding `line-less' galaxies into the analysis and the effect on the derived EW distribution.

The LIER spaxels in eLIER galaxies have on average lower EW(H$\alpha$) than LIER spaxels in cLIER galaxies. In particular, the EW(H$\alpha$) distribution for cLIERs (red in Fig. \ref{fig6.1}) shows a prominent tail at large EW, which is an effect of the contamination from star forming regions on approaching the LIER-star formation interface. On the other hand, even in eLIER galaxies a small number of localised regions display EW(H$\alpha$) $>$ 3\AA. These spaxels generally correspond to well-defined bisymmetric structures in the equivalent width maps resembling an integral sign (as observed also in the \textsc{SAURON} survey by \citealt{Sarzi2006, Sarzi2010}). Examples of these features in the MaNGA data are presented in Cheung et al. (\textit{submitted}).

The spaxels associated with LIER emission in SF galaxies (i.e. DIG, green in Fig. \ref{fig6.1}), show higher EW than other LIER regions. This can be interpreted as the result of the fact that the local stellar continuum is not the source of the ionising radiation, which instead comes from leakage of ionising photons from star forming regions elsewhere in the galaxy. Since most of these LIER regions are extra-planar, the continuum emission is weak, which translates into high EW.

\subsection{H$\alpha$ radial gradients}
\label{sec6.2}

In Fig. \ref{fig6.2} we show the radial gradients for the H$\alpha$ surface brightness (top panels) and EW(H$\alpha$) (bottom panels) as a function of deprojected galactocentric radius (in units of effective radius $\rm R_e$), for SF, eLIER and cLIER galaxies. 

For eLIER galaxies the H$\alpha$ surface brightness decreases steeply with radius, following the rapid drop of the stellar continuum. The fact that the line emission follows the continuum becomes clearer by looking at the radial gradient of the EW(H$\alpha$), which shows a nearly flat profile with narrow dispersion around a low value ($\sim 1.5$ \AA). Note, however, that the median EW(H$\alpha$) may be biased high because of the S/N cut imposed on line emission, as the MaNGA sensitivity limit corresponds roughly to EW(H$\alpha$) $\sim$ 1.0 \AA.

For SF galaxies the radial distribution of H$\alpha$ surface brightness shows a larger dispersion and decreases more slowly with galactocentric radius. The EW(H$\alpha$) has a much higher mean value than for eLIERs (typically $> 10$ \AA), with a relatively constant profile as a function of galactocentric radius, and with a weak, but significant drop toward the central region, associated with some contribution from the bulge to the central continuum emission. Since for SF galaxies the EW(H$\alpha$) is a good proxy of the specific SFR (the SFR per unit mass), we conclude that the mean specific SFR is only weakly dependent on galactocentric radius, although it decreases slightly in the central regions ($\rm R < 1.0 ~R_e$). This is in agreement with the observations of a spatially resolved relation between stellar mass surface density and SFR \citep{Cano-Diaz2016}. 

cLIER galaxies (central panels in Fig. \ref{fig6.2}) have more complex gradients. In the central region their EW(H$\alpha$) approaches the same low values observed in eLIERs, pointing towards a similar excitation mechanism in such regions, while the outer regions (where star formation occurs) have a higher EW(H$\alpha$), although generally slightly lower than in SF galaxy discs, suggesting some intrinsic differences in terms of age of the stellar population and mixture between ongoing and evolved stellar populations. The H$\alpha$ surface brightness has a complex profile resulting from a combination of steeply declining central profile (as in eLIERs) and an increase in flux at the radius where the LIER-star formation transition occurs. In the stacked profile presented in Fig. \ref{fig6.2} the effect is evident as a large increase in dispersion at $\rm 0.5 < R/R_{e} < 1.0$, due to the fact that the LIER-star formation transition occurs at different radii in different galaxies.

Several authors \citep{Sarzi2010, Yan2012, Singh2013} have investigated the radial profile of line emission in LIER galaxies as a means of testing the hypothesis of illumination from a central source. To compare with these studies we show here the radial profiles of  H$\alpha$ surface brightness ($\rm \Sigma_{H\alpha}$) for eLIER galaxies. We do not show the equivalent plot for cLIERs because the transition to SF regions at $\rm 0.5 < R/R_{e} < 1.0$ means that only the central profiles, which are more heavily affected by beam smearing, could be used for this test.

Fig. \ref{fig6.3}, left panel, shows the $\rm \Sigma_{H\alpha}$ gradients normalised to the central H$\alpha$ flux as a function of the deprojected distance from the galaxy centre for eLIER galaxies. The black solid line represents a $\rm 1/r^2$ intensity profile, convolved with the median MaNGA PSF, which is the expected surface brightness profile in case of photoionization by a central source (e.g. AGN) assuming a constant cloud number density and volume filling factor as a function of radius. Unfortunately cloud number densities and filling factors are not readily determined from the available data. Hence a line surface brightness profile shallower than $\rm 1/r^2$ provides only a weak indication regarding the origin of line emission in eLIER galaxies. Interestingly, although the individual galaxy profiles show significant scatter, our data is in general agreement with the $\rm 1/r^{1.28}$ profile found by indirect means by \cite{Yan2012}.
In Fig. \ref{fig6.3} we also show the EW(H$\alpha$) profiles for individual eLIER galaxies, demonstrating the flatness of the profiles on a galaxy by galaxy basis and with very few galaxies showing EW(H$\alpha$) $>$ 3 \AA. Galaxy {\tt 1-2117022}, which shows and unusually high average EW, is discussed in detail in Cheung et al. (\textit{submitted}).

\section{Line ratio gradients}
\label{sec7}

\begin{figure*} 
\includegraphics[width=0.9\textwidth, trim=0 170 0 180, clip]{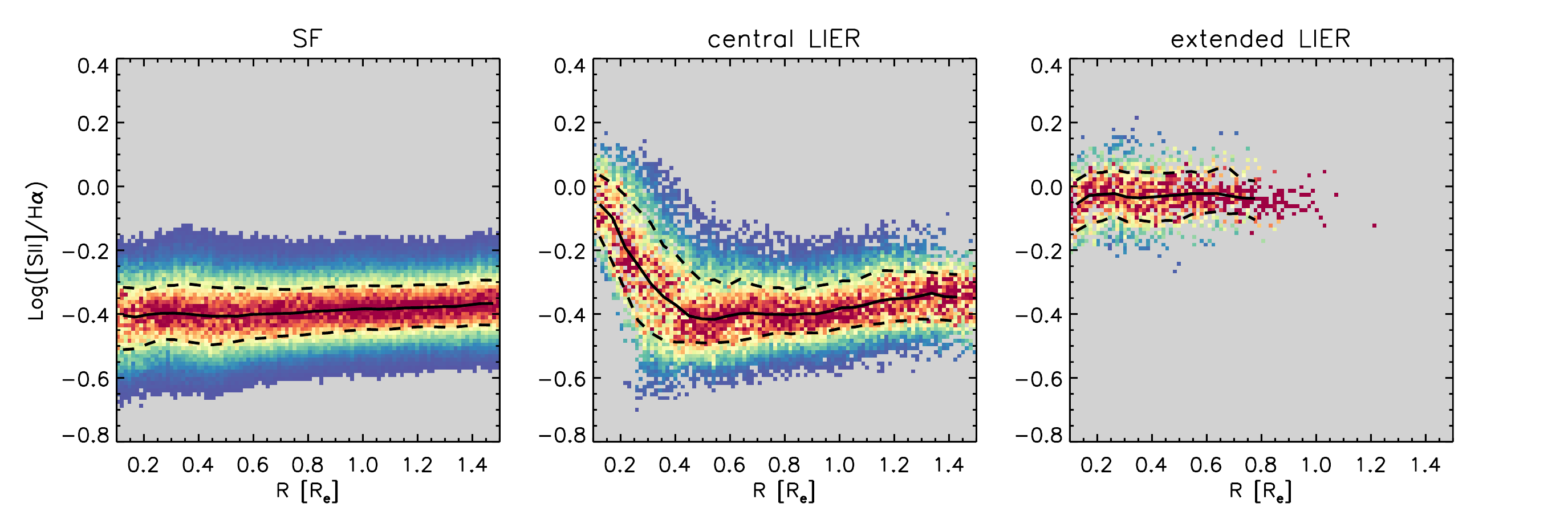}
\includegraphics[width=0.9\textwidth,  trim=0 170 0 180, clip]{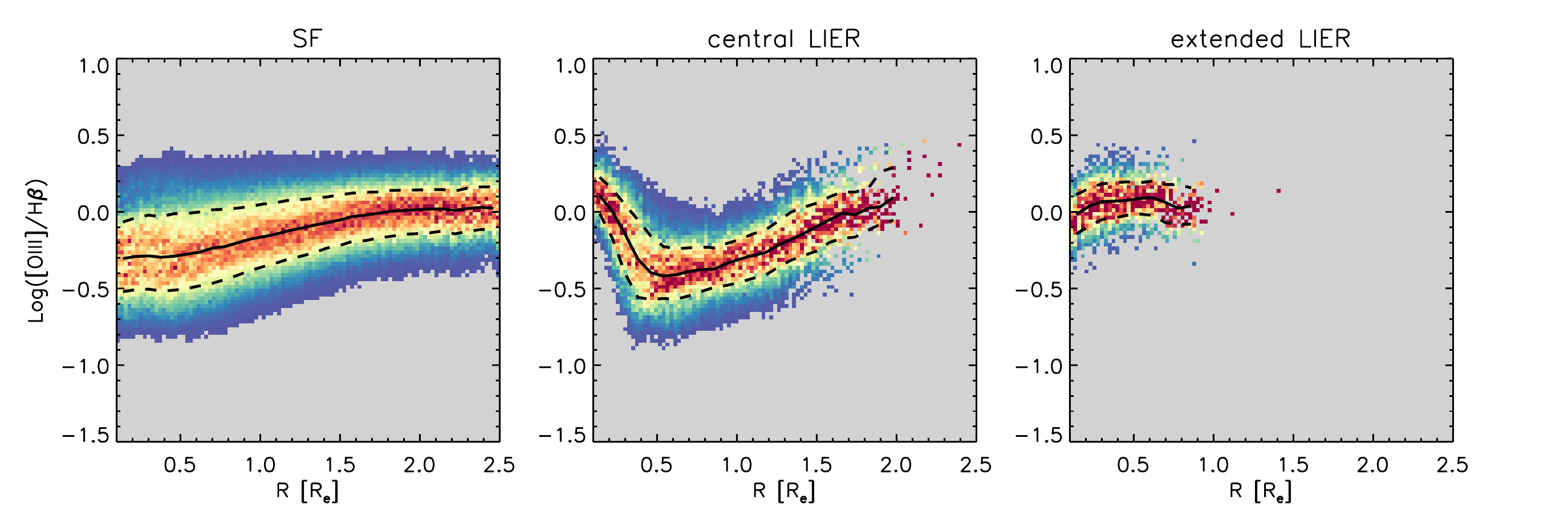}
\includegraphics[width=0.9\textwidth,  trim=0 170 0 180, clip]{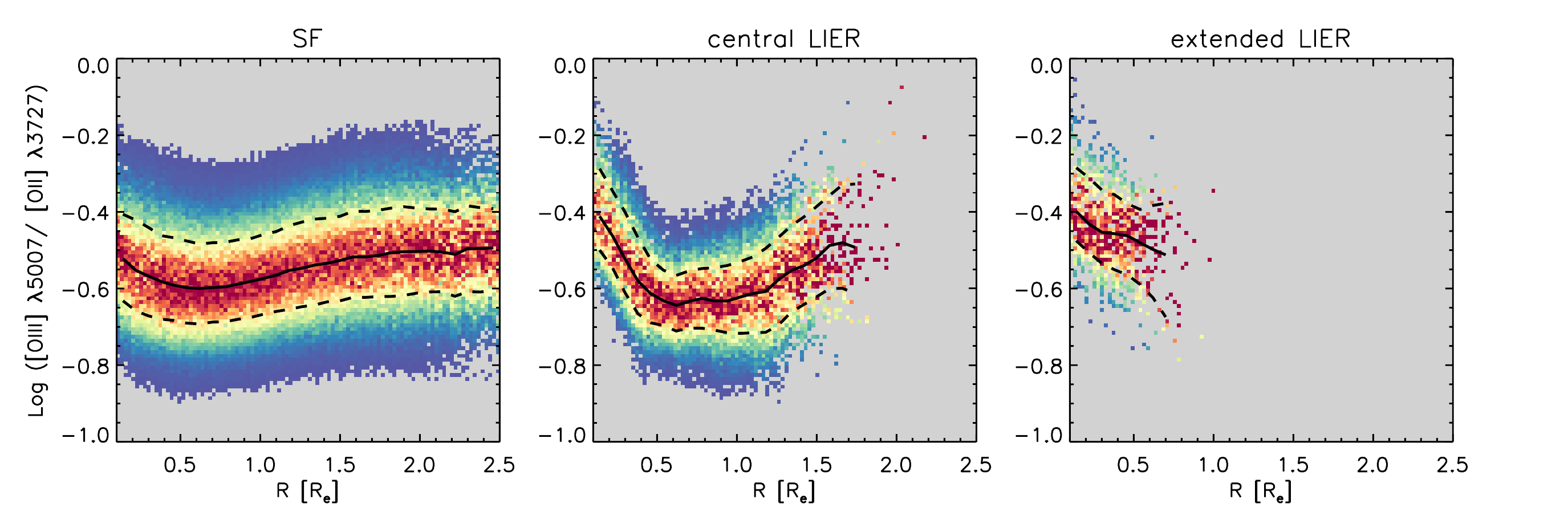}

\caption{Gradients of different diagnostic ratios as a function of deprojected radius in units of $\rm R_e$ for SF, cLIER and eLIER galaxies. From top to bottom: 
1) [OII]/H$\beta$, 2)[SII]/H$\alpha$ and 3)[OIII]/[OII]. The spaxels are binned in the 2D space and only bins with more than 4 spaxels are shown. At each radius the distributions are normalised so that the colour-coding represents the density of points at that radius. The red colour is fixed to represent the mean of the distribution at each radius. The black solid line represent the median at each radius, and the dashed lines represent the 25th and 75th percentiles.}
\label{fig7}
\end{figure*}

Line ratios can be used as probes of the fundamental parameters of the gas and the ionising radiation field: the gas electron density ($\rm n_e$), ionisation parameter (U), gas phase metallicity (Z) and hardness of the ionisation field ($\alpha$). 

Doublets of a single species in which the two levels have different collision strengths, like [OII]$\lambda\lambda$3726,28 or [SII]$\lambda\lambda$6716,31 are useful density tracers in the optical wavelength range. Note that, if the emission is clumpy, line emission effectively traces the density of the clumps and not the global average density.

The ionisation parameter (U) encodes information about the relative geometry of the gas and its ionising source. Considering a line-emitting cloud at a distance r from an ionising source, the ionisation parameter U(r) is defined as the dimensionless ratio between the number density of ionising photons and the electron density ($\rm n_e$)

\begin{equation}
\rm U(r) = Q/(4~\pi~r^2~n_e~c)
\end{equation}
where Q is the rate of ionising photons incident on the cloud. In a fully ionised medium, the electronic density is approximately the hydrogen gas density. Hence, the ionisation parameter can be seen as the dimensionless ratio between the number density of hydrogen-ionising photons and the number density of the total hydrogen gas (including ionised, neutral and molecular, as defined in \citealt{Osterbrock2006}, see also \citealt{Sanders2015}).

In photoionisation equilibrium the ionisation parameter is proportional to the ratio of the population in two ionisation states of the same element. Within the MaNGA wavelength range, the [OIII]$\lambda$5007/[OII]$\lambda\lambda$3726,28 ratio is a good proxy for the ionisation parameter, when calibrated through appropriate sets of photoionisation models which take into account the details of the ionising source. The hardness of the ionisation field ($\alpha$), which parametrizes the power law slope of the ionising spectrum per unit frequency, also affects the [OIII]$\lambda$5007/[OII]$\lambda\lambda$3726,28 ratio, by changing the ionisation structure of the nebula. This is particularly relevant in case of LIER excitation, as previous work \citep{Binette1994, Stasinska2008, Yan2012} has demonstrated that LIER emission is best reproduced by photoionisation by a harder radiation spectrum than that of young stars and a lower ionisation parameter than that of Seyfert AGN. 

The flux ratio of the [SII] doublet, [S II]$\lambda$6731/[S II]$\lambda$6717, constitutes as direct probe of the gas electron density, with the ratio expected to vary from $\sim$ 0.7, for $\rm n_e \sim 10^{1.5} cm^{-3}$ to $\rm \sim$ 2.0, for $\rm n_e \sim 10^4 cm^{-3}$ \citep{Osterbrock2006}. The sensitivity of this ratio saturates at the low electron density value for most regions in MaNGA galaxies (both SF and LIERs), causing the [SII] doublet ratio to appear artificially flat as a function of radius for all galaxy classes. The median [SII] ratio for SF galaxies is 0.725, corresponding to a density $n_e \sim 10^{1.5}$, close to the lower limit at which the line ratio saturates.


In Fig. \ref{fig7} we show the gradients of [SII]/H$\alpha$ for different galaxy classes. Since [SII] is a low ionisation line, emitted primarily in the partially-ionised region of the gas cloud, [SII]/H$\alpha$ is a good tracer or the importance of low-ionisation species and thus provides a good separation between star formation and LIER ionisation. The power of this diagnostic is demonstrated in Fig. \ref{fig7} by the effective bimodality between the SF (log([SII]/H$\alpha$) $\sim$ -0.4) and LIER locus (log([SII]/H$\alpha$) $\sim$ -0.05).

\begin{figure} 
\includegraphics[width=0.45\textwidth, trim=20 45 60 60, clip]{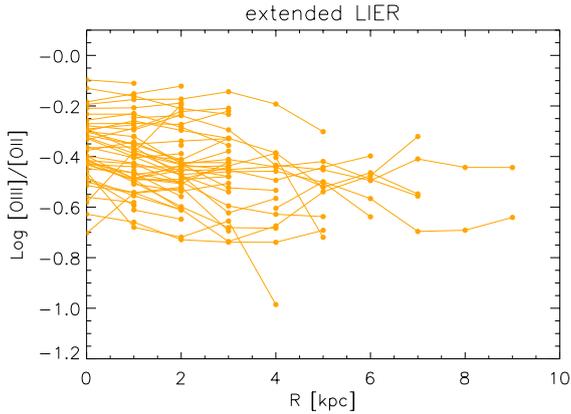}
\caption{Radial gradient of the [OIII]$\lambda$5007/[OII]$\lambda\lambda$3727-29 ratio corrected for extinction as a function of deprojected radius for individual eLIER galaxies.}
\label{fig7.1}
\end{figure}

Both the [OIII]/[OII] and [OIII]/H$\beta$ line ratios are sensitive to the ionisation parameter and the hardness of the ionisation field. [OIII]/H$\beta$ has a more pronounced metallicity dependence than [OIII]/[OII], which is roughly proportional to U, for both H\textsc{ii} regions \citep{Diaz2000} and harder AGN or old stellar population spectrum \citep[e.g.][]{Yan2012}. In order to account for the large wavelength difference between [OII]$\lambda$5007 and [OII]$\lambda$3726,29 we perform an extinction correction based on the Balmer decrement to measure the [OIII]/[OII] ratio. Fig. \ref{fig7} shows the gradients in  [OIII]/[OII] and [OIII]/H$\beta$ for SF, cLIER and eLIER galaxies. By selection, regions dominated by LIER ionisation in both cLIERs and eLIERs show common line ratios, characterised by higher [OIII]/[OII] and [OIII]/H$\beta$  than SF galaxies. Remarkably, for eLIERs the gradient in [OIII]/H$\beta$ is nearly flat and the [OIII]/[OII] gradient is only weakly decreasing as a function of radius ($\sim 0.2$ dex decrease over 1 $\rm R_e$). In Fig. \ref{fig7.1} we show the [OIII]/[OII] for individual eLIER galaxies as a function of physical radius, demonstrating that gradients for individual galaxies are approximately flat for several kpc.

The ionisation parameter, being directly proportional to the ionising photon flux, constitutes a constrain on the geometry of the excitation source. Assuming the electron density to be approximately constant over the range of radii covered by MaNGA, the ionisation parameter should go down as $\rm 1/r^2$ in the same radial range in the case of point source illumination (e.g. a central AGN). However, several caveats apply this arguments. Firstly, the electron density is likely to decrease with radius, although most likely following a more shallow power law than $\rm r^{-2}$ \citep{Sarzi2010}. 

Moreover, in nearby resolved narrow line regions of Seyfert galaxies the ionisation parameter is generally found to be roughly constant as a function of radius, and also to be remarkably constant across different sources \citep[e.g.][]{Veilleux1987, Veron-Cetty2000}. A variety of models have been proposed to explain these observations, the most successful of which are dusty radiation pressure-dominated AGN models \citep{Dopita2002, Groves2004a, Groves2004b}. In these models, for ionisation parameters greater than U $\sim \alpha_B/ (c~\kappa_d) \sim10^{-2}$, where $\alpha_B$ is the case B recombination coefficient for hydrogen and $\kappa_d$ is the dust effective opactiy, the narrow line region is radiation pressure-dominated \citep{Dopita2002}. In this regime the ionisation parameter of the line-emitting clouds remains roughly constant as a function of radius. 

The ionisation parameters implied by the observed [OIII]/[OII] ratio in eLIER is U $\sim 10^{-4}-10^{-3}$ \citep{Binette1994, Stasinska2008}, which is 1-2 orders of magnitude lower than the critical value of the ionisation parameter which would lead to radiation-pressure domination. Thus even for classical LINER-type AGN, radiation pressure is predicted to be negligible. We conclude that the observed flat profiles in the ionisation parameter-sensitive line ratio [OIII]/[OII] are not consistent with a low-ionisation parameter, weak AGN scenario.

Overall, the MaNGA data are in good agreement with previous studies of gradients of line ratios in elliptical galaxies, which generally demonstrates very shallow or flat line profiles outside the central 100 pc \citep{Annibali2010, Papaderos2013, Sarzi2010, Yan2012}.

\section{Age-sensitive stellar population indices}
\label{sec8}

\subsection{Age-sensitive indices distributions}

\begin{figure*} 
\includegraphics[width=0.49\textwidth, trim=80 90 80 150, clip]{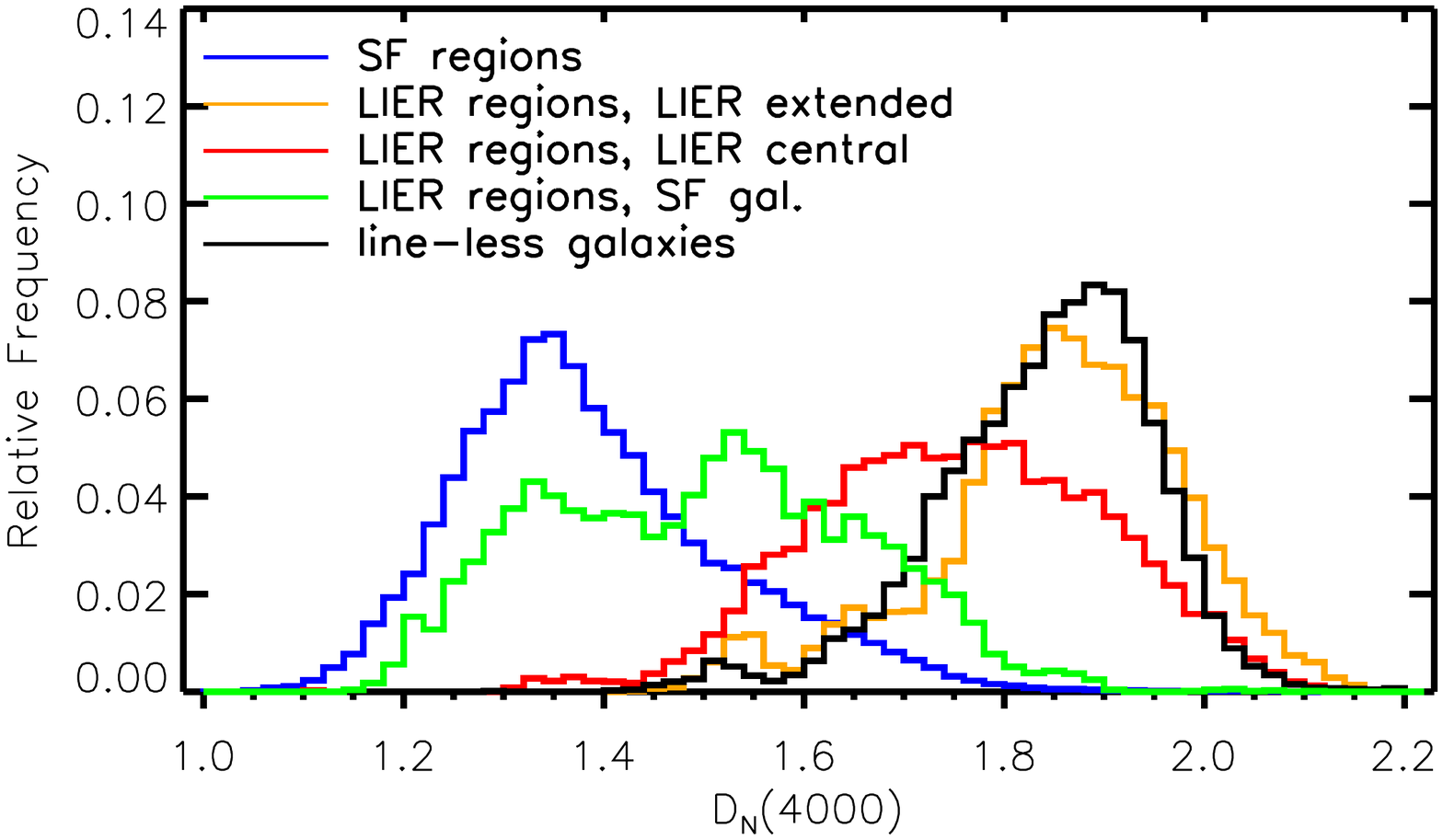}
\includegraphics[width=0.49\textwidth, trim=80 90 80 150, clip]{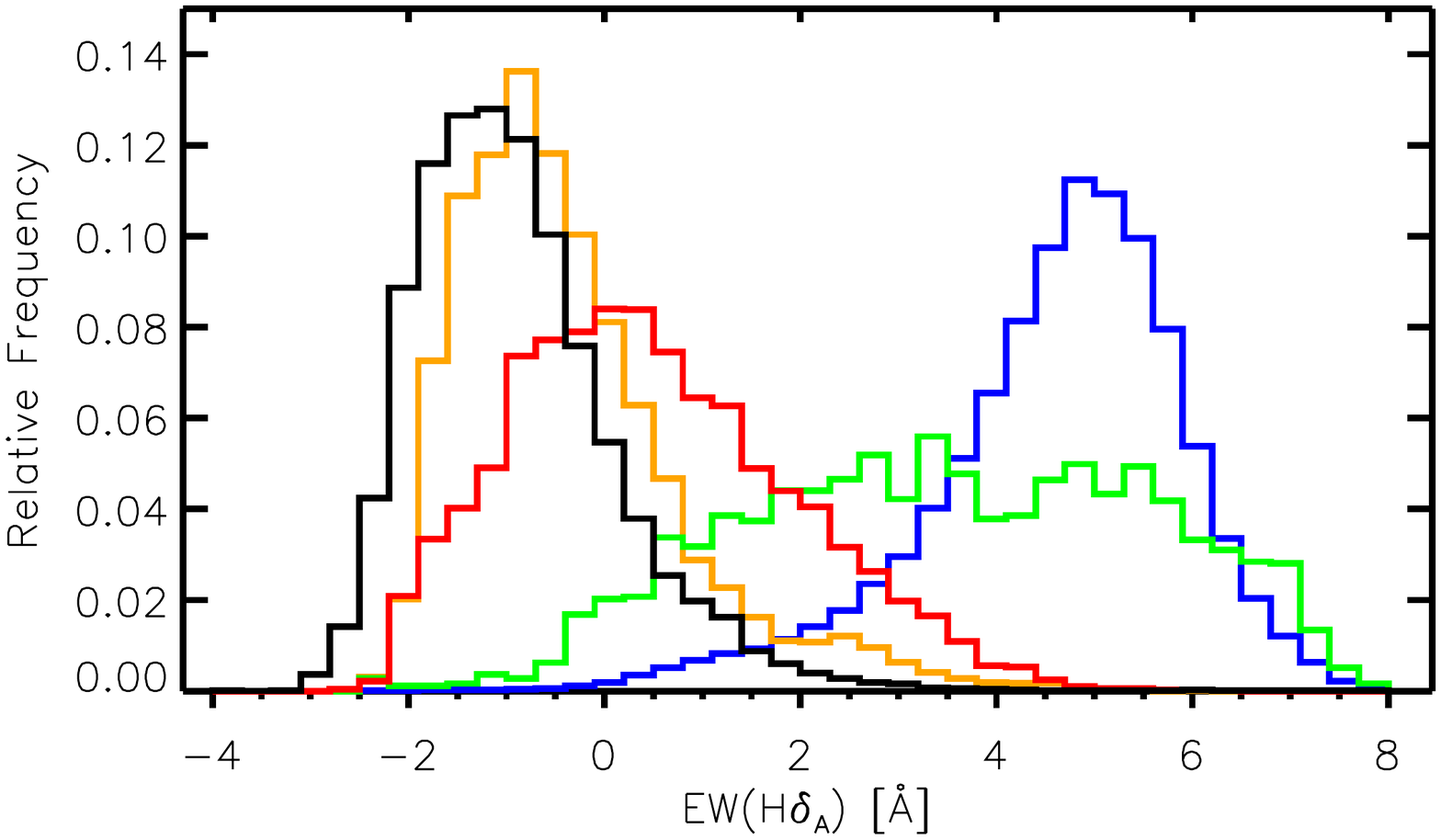}
\caption{Histograms of the distribution of $\rm D_N(4000)$ and  $\rm EW(H\delta_A)$ in MaNGA spaxels. Spaxels of different classes are colour-coded differently following the legend in the left panel. The blue histograms correspond  to spaxels classified as SF in SF galaxies, while SF spaxels in cLIER galaxies are plotted in violet. LIER spaxels are subdivided in three classes, depending on the galaxy they belong to: eLIER (orange), cLIER (red). Finally, spaxels were line emission is not detected (line-less) are plotted in black.}
\label{fig8.1}

\includegraphics[width=\textwidth,  trim=0 200 0 200, clip]{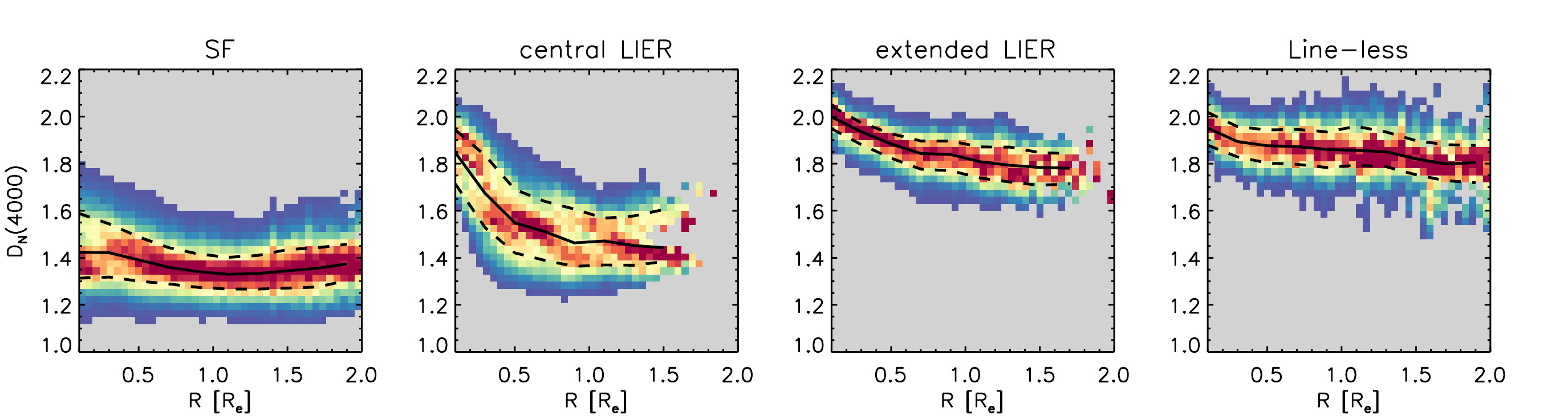}
\includegraphics[width=\textwidth,  trim=0 200 0 220, clip]{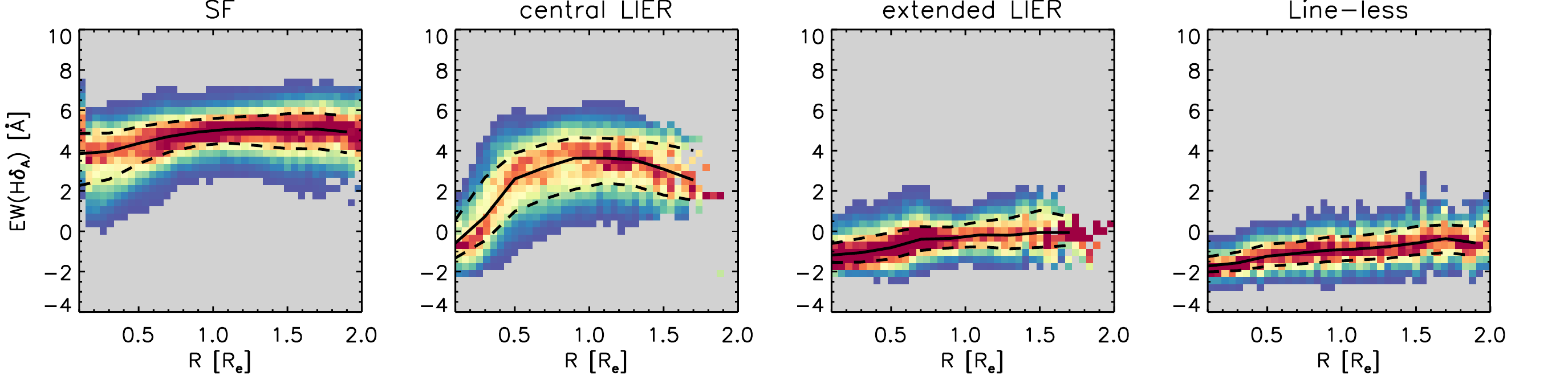}
\caption{Gradients of $\rm D_N(4000)$ (top) $\rm EW(H\delta_A)$ (bottom) and as a function of deprojected effective radius for MaNGA galaxies split according to the emission-line based classification introduced in this work. The spaxels are binned in the 2D space and only bins with more than 10 spaxels are shown. At each radius the distributions are normalised so that the colour-coding represents the density of points at that radius, with the red colour fixed to represent the mean of the distribution at each radius. The black solid line represent the median at each radius, and the dashed lines represent the 25th and 75th percentiles.}
\label{fig8.2}

\end{figure*}

In Sec. \ref{sec3.2} we argued that LIER emission is associated with old stellar population, as traced by the stellar population indices $\rm D_N(4000)$ and $\rm EW(H\delta_A)$ in legacy SDSS spectroscopy as well as in the spatially resolved MaNGA data. In this section we further study differences in the distribution and radial gradients of these age sensitive stellar population indices in the MaNGA data. We split the MaNGA spaxels in the four classes defined in Sec. \ref{sec6.1}: 1) SF spaxels, 2) LIER spaxels in cLIER galaxies, 3) LIER spaxels in eLIER galaxies and 4) LIER spaxels in SF galaxies. Moreover, in this section we additionally consider spaxels with reliable continuum but undetected line emission in line-less galaxies.

Fig. \ref{fig8.1} shows the histogram distribution of $\rm D_N(4000)$ and $\rm EW(H\delta_A)$ for the MaNGA spaxels divided in the classes defined above. SF regions in SF galaxies have stellar population indices consistent with young stellar populations ($\rm D_N(4000) \sim 1.3$ and  $\rm EW(H\delta_A) \sim 5$). On the opposite end of the spectrum, LIER spaxels in eLIER galaxies and line-less spaxels are characterised by old stellar populations. In fact, the presence (eLIER) or absence (line-less) of line emission does not have any effect on the stellar populations in these systems. I.e. the stellar populations properties of eLIER and line-less galaxies are fully consistent with each other. 
LIER regions in cLIER galaxies show $\rm D_N(4000)$ and $\rm EW(H\delta_A)$ distributions which are intermediate between SF and eLIER/line-less galaxies. Intermediate values of these indices require careful modelling since in these regime both indices are sensitive to both metallicity and stellar age. 
Nonetheless the observation are consistent with cLIER not having any young stars (age $<$ 1 Gyr) in their LIER regions.

As anticipated in Sec. \ref{sec4.1} (and Fig. \ref{fig4.1}) the only LIER regions that host young stars ($\rm D_N(4000) <$ 1.5) correspond to LIER regions in SF galaxies, associated with DIG in both extra-planar and inter-arm regions in spirals.
In order to put this analysis into context it is important to remember that the different galaxy classes have different stellar mass distributions. Because of the well-known downsizing effect, lower mass galaxies are also younger and thus it is possible that SF galaxies appear on average younger than cLIERs because cLIERs occur more frequently at higher mass (see Fig. \ref{fig5.2}). We have therefore divided the sample in stellar mass bins and checked that the trends presented in Fig. \ref{fig8.1} are preserved. This is indeed the case because, while on average the stellar populations of all galaxy classes become older with increasing stellar mass, the relative stellar indices differences between classes of spaxels are preserved in each stellar mass bin ($\rm 10 < log(M_\star/M_\odot) <11.5$). At the extremes of the stellar mass distribution we do not have enough galaxies of each class to be able to perform this test.

\subsection{Age radial gradients}

Fig. \ref{fig8.2} shows gradients of $\rm D_N(4000)$ and $\rm EW(H\delta_A)$ for MaNGA galaxies of different classes. A detailed study of the dependence of stellar population properties gradients on other galactic properties will be presented in future work. Here we only comment on the qualitative differences between gradients in galaxies classified according to our emission-line based classification. 

\begin{enumerate}
\item{SF galaxies have flat $\rm D_N(4000)$ and $\rm EW(H\delta_A)$, consistently with the fact that young stellar populations are present at all radii. Older stellar populations are found for a subset of the most massive galaxies in the nuclear regions (galaxies with $\rm log(M_\star/M_\odot ) > 10.5$, corresponding to an increase in dispersion for $\rm r < 0.5 \ R_e$)}, which can be associated with an older bulge component.

\item{eLIER and line-less galaxies show slowly radially decreasing $\rm D_N(4000)$ and increasing $\rm EW(H\delta_A)$, consistent with previous study of age profiles in early type galaxies \citep{Wilkinson2015, Li2015}. However the observed gradient in the two stellar indices is due to a combination of both metallicity and age effects and a combined analysis is needed to reliably disentangle the two.}

\item{cLIER galaxies show a clear transition between old stellar populations in the inner (LIER) region and young stellar populations in the outer (SF) regions.}
\end{enumerate}

Goddard et al. ({\it in prep.}) make use of full spectral fitting \citep{Heavens2000, Panter2004, CidFernandes2005, Gallazzi2006, Sanchez-Blazquez2011, Wilkinson2015, McDermid2015a} to derive age and metallicity of the stellar population, thus largely mitigating the age-metallicity degeneracy. However the bimodality observed in stellar population between `young' ($\rm D_N(4000) <$ 1.5) and `old' ($\rm D_N(4000) >$ 1.5) is very robust to metallicity effects \citep{Wilkinson2015}. 

The gradients in stellar indices discussed above are diluted due to the combination, within each galaxy class, of galaxies with different properties (masses, environments etc.). However, given the well-defined differences in stellar indices gradients between the different galaxy classes, an emission-line based classification appears to capture a sharp distinction between young (SF), transitioning (cLIER) and quiescent (eLIER and line-less) galaxies, with other properties (like stellar mass) representing a smooth variation in addition to the basic trend presented here.


\section{Discussion}
\label{dis}

\subsection{Energetics: powering the observed line emission in LIER galaxies}
\label{sec8.1}
A fundamental finding of this work is that typical weak AGN do not emit sufficient energy in ionising photons to explain the observed line emission in LIER galaxies (see Sec. \ref{sec3.5}). pAGB stars, on the other hand, represent a diffuse source of ionising photons, which must be present in galaxies with old stellar populations. Although further work is needed to reduce the uncertainties in current modelling of the late stages of stellar evolution \citep[e.g.][]{O'Connell1999}, considering the uncertainty in the modelling parameters, H$\alpha$ equivalent width of 0.5 - 2.0 \AA\ can be generated by the pAGB stellar component. The tight correlation between the observed continuum and line emission in LIERs, and the fact that the EW(H$\alpha$) observed in LIER regions lies in the range predicted by the models is strong evidence in favour of the stellar hypothesis for LIER emission.

This conclusion should not be considered at odds with studies of truly nuclear LINERs on $\sim$ 100 parsec scales. For example \cite{Ho2008} estimates that on the scales of 100-200 pc covered by the Palomar spectroscopic survey, pAGB stars can only account for $\sim$ 30-40\% of the emission observed in LI(N)ER nuclei. On these small scales a weak AGN could indeed be the main source of ionising photons (see \ref{sec3.5}). However, at the resolution of surveys like MaNGA (or within the $3''$ SDSS fibre), a possible LINER AGN contribution is generally subdominant beyond the central resolution element.

Strong gradients and localised features in either the equivalent width and/or diagnostic line ratios are generally not expected if LIER emission is attributed to diffuse stellar sources. Indeed, the flat gradients and low observed EW(H$\alpha$) are consistent with this scenario. However, even in eLIER galaxies, a small number of regions lie in the upper tail of the EW(H$\alpha$) distribution with EW(H$\alpha$) = 3-6 \AA\ (Fig. \ref{fig6.2}). Interestingly, these features generally correspond to localised features, often of bisymmetric morphology (`integral sign', \citealt{Sarzi2006}, Cheung et al., \textit{submitted}). The self-similarity of these structures points towards a likely common origin. A possible explanation lies in the relative geometry of the ionising stellar sources and the gas cloud absorbers. If the gas clouds are not randomly distributed around the stars but preferentially concentrate along coherent structures on kpc scales (discs, ionisation cones, etc.), then diffuse ionisation sources like pAGB stars can be sufficient to explain the higher observed equivalent widths. Non-uniform dust concentration might also play a role in defining these structures. 

On the other hand, shocks provide a natural mechanism to generate localised features in line emission maps. A prediction of the shock scenario is that localised equivalent width enhancements should be accompanied by a systematic change in line ratios and, possibly, by an increase in the gas velocity dispersion due to unresolved velocity gradients. 

Finally, other sources of ionising radiation, such as low-mass X-ray binaries and extreme horizontal branch stars, have been found to be a subdominant contribution with respect to pAGB stars by at least one order of magnitude \citep{Sarzi2010, Yan2012}. Their contribution to the relevant energy budget is thus overall negligible unless the number of pAGB stars has been grossly over-predicted by models. 

An interesting note on this topic is represented by the fact that \textit{Hubble Space Telescope (HST)} imaging of nearby LIERs (like M31, \citealt{Ciardullo1988, Heckman1996}, which in a MaNGA-like observation would appear as a cLIER, or the nearby elliptical galaxy M32) struggles to account for the number of UV bright stars predicted by stellar population models \citep{Brown1998, Brown2008, Rosenfield2012}. At the moment it is not clear whether these inconsistencies are due to observational bias, limitations of the theoretical models or both. Clearly more work is needed in this area to be able to conclusively associate the LIER emission to the UV stellar sources powering it.

\begin{figure*} 
\includegraphics[width=\textwidth, trim=0 30 0 0, clip]{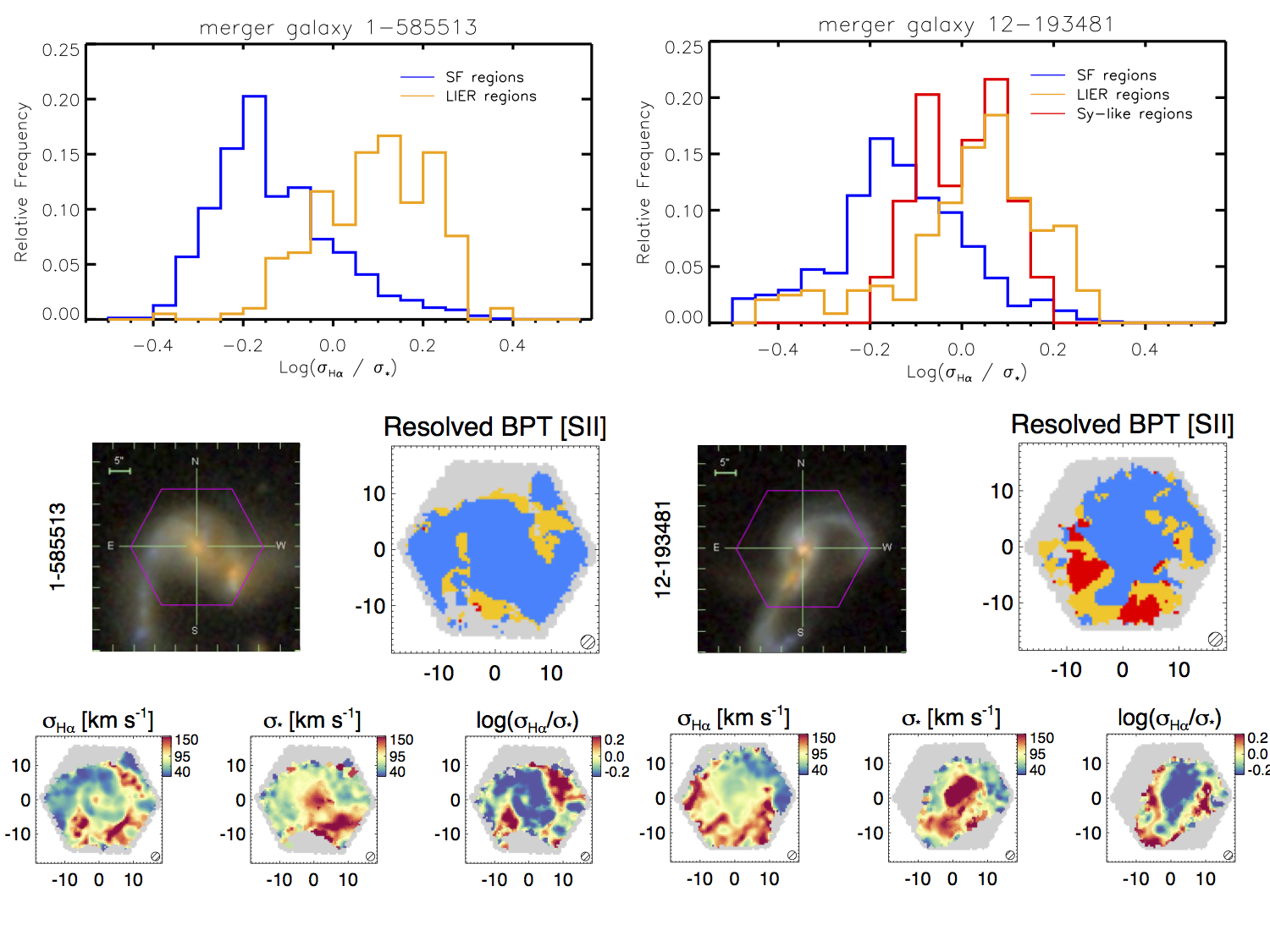}
\caption{An illustration of the properties of LIER-like emission in two major mergers. The top panels show the $\rm log(\sigma_{H\alpha}/\sigma_\star$) ratio for the SF (blue), LIER-like (orange) and Sy-like (red) spaxels, demonstrating the fact the dispersion in LIER-like regions in mergers is dominated by non-gravitational motions. In the second row we show the $g-r-i$ SDSS composite image and the resolved BPT diagram, based on the position of galactic regions in the [SII] BPT diagram, as in Fig. \protect\ref{fig5}. In the third row we show the H$\alpha$ and stellar velocity dispersion maps, together with a map of the $\rm log(\sigma_{H\alpha}/\sigma_\star$) ratio.}
\label{fig8.33}
\end{figure*}

\subsection{Local ionisation conditions determine the state of the LIER gas}

The question of the relative distance and geometry of ionising sources and line emitting gas in LIER galaxies is crucial to the correct interpretation of the MaNGA observations presented in this work. Both \cite{Sarzi2010} and \cite{Yan2012} have presented a simple model for emission by diffuse stellar sources in elliptical galaxies, based on the assumption of spherical symmetry and random positions of the line emitting gas clouds with respect to the stars. The main power of these models lies in their ability to demonstrate that shallow profiles in EW and logU are naturally produced by a wide range of parameters when diffuse sources are assumed to the the source of ionising photons. 

However, these models generally assume an infinite mean free path for the ionising photons, thus producing line emission and EW profiles which are not constant as a function of radius. A smaller mean free path for the ionising photons, due to a smaller gas-stars distance, leads to shallower profiles of line emission, as the line emission becomes more tightly coupled to the `local' ionising sources. The observed remarkable flatness of the EW(H$\alpha$) profile and other line ratios diagnostics support the scenario where the `local' continuum is directly powering the observed line emission within the resolution element considered (which is approximately kpc size for the MaNGA data).

Indeed, assuming that a single pAGB star emits the reference value of $\rm 10^{47}$ ionising photons per second and an electron density of $\rm 10^2~ cm^{-3}$, in agreement with the measured [SII] double ratio, for $\rm logU = 10^{-4}$ (consistent with the observed values of [OIII]/[OII] in  LIERs, as discussed in \citealt{Binette1994} and \citealt{Stasinska2008}) the mean separation between ionising source and gas cloud must be of order of a few parsecs, well below our kpc-scale resolution. The small separation between gas and pAGB stars is naturally explained if the gas is assumed to be the result of stellar mass loss from the pAGB stars in earlier stages of their evolution. In general, gaseous material from stellar mass loss would be expected to be co-spatial with the stellar continuum on the kpc scales, thus further justifying the `locality' assumption when calculating the EW(H$\alpha$).

In this scenario, geometry is also a likely explanation for EW(H$\alpha$) lower than 3 \AA, which can be explained by relaxing the assumption that all photons are absorbed `locally'. In particular, in galaxies devoid of gas is possible that a fraction of the ionising photons escape the galaxy altogether. In fact, in the pAGB scenario is it harder to justify the absence of line emission than its presence. In \textit{Paper II} we present a discussion of the possible reasons for lack of line emission in line-less galaxies.


\subsection{The case for shock excitation in interacting/merging galaxies}
\label{sec8.3}

Shocks are the results of several astrophysical phenomena likely to occur in galaxies including supernovae explosions, AGN winds and the interaction of a jet with the ISM. It is well known that the BPT diagram does not represent a good test for shock ionisation, as shock models cover larger areas of both the LIER and SF sequences \citep{Allen2008,Rich2010, Rich2011, Ho2014, Alatalo2016}. Moreover, just like for the case of a weak AGN, the presence of shocks does not imply that they must be the main source of ionising photons in galaxies.

A correlation between line ratio and velocity dispersion has often been considered a good test for shock ionisation \citep{Monreal-Ibero2006, Rich2011, Rich2014}. However in the case of LIER galaxies this test is of difficult interpretation since, inside a spheroidal component, the velocity dispersion of the stars (and possibly of the gas) has a radial dependence due entirely to gravitational virial motions. A potentially more promising way for identifying shocks is looking for a localised increase in the velocity dispersion of the gas compared to that of the stars. Since the gas is generally dynamically colder than the stars, due to its ability to dissipate non circular motions, and the fact that the thermal broadening of the gas at $\rm 10^4 K$ is a negligible contribution to its velocity dispersion, a large increase in the gas velocity dispersion compared to the stellar dispersion is a possible signature of shocks. A study of the stellar and gas velocity dispersion of the MaNGA galaxies is presented in \textit{Paper II}, demonstrating that the gas velocity dispersion is statistically comparable or lower than the stellar velocity dispersion, even in eLIER galaxies.

In this study we only discuss the potential of the method in isolating shocked regions in the few interacting galaxies in the current MaNGA sample. As briefly discussed in Sec. \ref{sec5.1}, gas rich mergers show localised regions of LIER-like (and even Seyfert-like) emission. In Fig. \ref{fig8.33} we show the histograms of $\rm \log(\sigma_{H\alpha}/\sigma_\star)$  for SF spaxels (blue), LIER-like (orange) and Sy-like (red) in two major mergers (MaNGA-ID {\tt 1-585513} and {\tt 12-193481}). LIER emission in merging systems shows systematically higher gas velocity dispersion compared to the stars, and, as demonstrated by the maps of velocity dispersion of gas and stars, this increase is highly localised in regions of LIER and Sy-like emission. These observations support the presence of LIER-like shock excitation in merging systems, consistently with the strong LIER-like emission observed in ULIRGs, \citep{Monreal-Ibero2006, Alonso-Herrero2010}, most of which are strongly starbursting merging systems, large-scale outflows \citep{Lipari2004} and regions at the jet-ISM interface \citep{Cecil2000}. We note, however, that in the low-z Universe merging systems are extremely rare and constitute a statistically negligible part of the overall galaxy population, although they may be more commonly observed at higher redshift. 

In summary, there is no evidence for an enhancement in the velocity dispersion of the gas with respect to the stars in normal cLIER and eLIER galaxies, leading to the conclusion that no unresolved velocity gradients due to shocks are necessary to explain the dynamical state of the gas. However, LIER emission in gas rich mergers is accompanied by a large increase in the gas velocity dispersion, leading to the conclusion that shock excitation plays a role in these rare systems.

\section{Conclusions}
\label{sum}

In this paper we used the BPT diagnostic diagrams ([NII]/H$\alpha$ and [SII]/H$\alpha$ versus [OIII]/H$\beta$) to study the spatially resolved excitation properties of a large sample of 646 galaxies with integral field spectroscopic data from the first 8 months of operations of the SDSS-IV MaNGA survey. This galaxy sample is roughly flat in stellar mass in the range $\rm 10^9 - 10^{12} M_\odot$ and spans both the red sequence and the blue cloud. Within each galaxy we make use of the [SII] BPT diagram to map the spatial extent and morphology of star formation, low ionisation emission line regions (LIERs) and Seyfert ionisation. We observe that, excluding mergers and BPT-classified Seyfert galaxies, normal galaxies are characterised by remarkably regular excitation morphologies. We suggest the following classification scheme to be applicable to the majority of the galaxy population.

\begin{enumerate}
\item{\textit{Line-less galaxies}: No detected line emission (EW(H$\alpha)<$ 1.0 \AA\ within 1.0 $\rm R_e$).}
\item{\textit{Extended LIER galaxies (eLIER)}: galaxies dominated by LIER emission at all radii where emission lines are detected. No evidence for any star forming regions.}
\item{\textit{Central LIER galaxies (cLIER)}: galaxies where LIER emission is resolved but spatially located in the central
regions, while ionisation from star formation dominates at larger galactocentric distances.}
\item{\textit{Star forming galaxies (SF)}: galaxies dominated by star formation in the central regions and at
all radii within the galaxy disc. LIER emission in these galaxies is sometimes observed in edge-on systems at large distances from the disc midplane and/or in inter-arm regions.}
\end{enumerate}

Focusing on the physical conditions of LIER regions within galaxies we have shown that:

\begin{itemize}
\item{
The extra-planar and inter-arm LIER emission in SF galaxies can be identified as diffuse ionised gas, an ionised gas component observed in both the Milky Way and other nearby external galaxies. The source of ionisation of the diffuse ionised gas is still unclear, however it is likely that escaping hardened radiation from star forming regions contributes to the ionisation budget, with a possible contribution from ionising photons from old stellar populations (Zhang et al., \textit{in prep.}). This hypothesis is consistent with the fact that LIER emission in SF galaxies lies close to the star formation-LIER demarcation line in the BPT diagram, implying a softer ionisation field that in true LIER galaxies, and a higher H$\alpha$ equivalent width, supporting ionisation from a non-local source (Sec. \ref{sec4}).}

\item{
cLIER and eLIER galaxies display equivalent LIER-like line ratios (Sec. \ref{sec7}) and are characterised by absence of young stellar populations (as traced by stellar indices like $\rm D_N(4000)$ and $\rm H\delta_A$, Sec. \ref{sec4.1} and \ref{sec8}) and low equivalent widths of H$\alpha$ emission (EW(H$\alpha$) $<$ 3 \AA, Sec. \ref{sec6}). The line emission follows closely the continuum surface brightness, or equivalently the EW(H$\alpha$) is nearly constant throughout the LIER regions, strongly suggesting the need for a diffuse, local ionising source to power the LIER line emission.}

\item{
Post asymptotic giant branch (pAGB) stars have been shown to produce the required hard ionising spectrum necessary to excite LIER emission \citep{Binette1994, Stasinska2008} and are thus the ideal candidate to power LIER emission in eLIER and cLIER galaxies. Stellar evolution models which included the pAGB phase demonstrate that the pAGB hypothesis is energetically viable for all LIER galaxies as the observed EW(H$\alpha$) lie in the same range as model predictions (0.5 - 3.0 \AA, Sec. \ref{sec3.2}). Although weak AGN may appear as true LINERs when observed on scales of 100 pc or less, a typical low luminosity AGN does not emit enough energy to be the dominant contributor to the ionising flux on the kpc scales probed by the 3$''$ SDSS fibre (from the legacy SDSS, $\rm <z> \sim$ 0.1) or by MaNGA ($\rm <z> \sim$ 0.03).}

\item{
Focusing on eLIER galaxies, where there is no contamination from star formation at large radii, we observe the surface brightness profiles of H$\alpha$ emission to be systematically shallower than $\rm 1/r^2$ (Sec. \ref{sec6}). Line ratios sensitive to the ionisation parameter ([OIII]$\lambda$5007/[OII]$\lambda$3727 and partially [OIII]$\lambda$5007/H$\beta$) show flat or very shallow profiles over radial scales of tens of kpc (Sec. \ref{sec7}). The fact that the radial gradient of the ionisation parameter does not follow an inverse square law, as expected for a constant electron density profile and illumination by a central source, further rules out AGN as sources for the extended LIER emission. This evidence is stronger than the observed shallow surface brightness of line emission, since line emission depends on the radial gradients of the gas clouds effective areas and volume filling factors, which are effectively unknown. }

\item{Shocks might play a role in the ionisation state of localised features observed in EW maps (Cheung et al., \textit{submitted}) and in merging/interacting galaxies (Sec. \ref{sec8.3}), but no evidence for shock excitation is found in the bulk population of cLIER and eLIER galaxies.}

\end{itemize}

Our results are in line with recent work on LIER emission in red galaxies \citep{Sarzi2010, Yan2012, Singh2013}, demonstrating that the line flux observed in the SDSS 3$''$ fibre is not related to AGN photoionisation. 
We suggest that, since LIER galaxies represent a significant fraction of all galaxies (up to 30\% in the stellar mass range $\rm 10^{10.5}-10^{11.5} ~M_\odot$) and are much more numerous that Seyfert nuclei, previous work studying the AGN census in SDSS may be in need of revision. Moreover, assuming pAGB stars are sources for LIER ionisation, LIER emission acquires a new role in the context of galaxy evolution as a probe of the properties of the ISM in regions where star formation has ceased. Intriguingly, the spatially resolved MaNGA spectroscopy demonstrates that these `retired' LIER regions do not only occur in classical early type galaxies (eLIERs), but also in the central regions of spirals (cLIERs). In other words, the observed bimodality between `star-forming' and `quiescent' (including LIERs) does not only apply to the integrated properties of galaxies but also to the sub-components within them. In a companion paper (Belfiore et al., \textit{in prep.}) we explore this idea further by studying both integrated and resolved properties of LIER galaxies with the aim of placing them within the framework of the evolving galaxy population.


\section*{Acknowledgements}
\begin{small}
F.B., R.M and K.M acknowledge funding from the United Kingdom Science and Technology Facilities Council (STFC). A.R-L acknowledges partial support from the DIULS regular project PR15143. M.B was supported by NSF/AST-1517006. K.B was supported by World Premier International Research Centre Initiative (WPI Initiative), MEXT, Japan and by JSPS KAKENHI Grant Number 15K17603. AW acknowledges support from a Leverhulme Early Career Fellowship. The authors are grateful to the referee for a thorough analysis of this paper which improved the clarity of the work; to L. Coccato for help with the MaNGA spectral fitting; to M. Blanton for developing and maintaining the NASA-Sloan Atlas; to M. Cappellari for sharing his IDL routines; to M. Blanton, M. Cappellari, D. Wake and M. Auger for helpful comments; to the members of the SDSS-IV MaNGA collaboration for support towards this project. This work makes use of data from SDSS-I-II and IV.

Funding for SDSS-I-II and SDSS-IV has been provided by the Alfred P.~Sloan Foundation and Participating Institutions. Additional funding for SDSS-II comes the National Science Foundation, the U.S. Department of Energy, the National Aeronautics and Space Administration, the Japanese Monbukagakusho, the Max Planck Society, and the Higher Education Funding Council for England.  Additional funding towards SDSS-IV has been provided by the U.S. Department of Energy Office of Science. SDSS-IV acknowledges support and resources from the Centre for High-Performance Computing at the University of Utah. The SDSS web site is {\tt www.sdss.org}.

The participating Institution in SDSS-II include the American Museum of Natural History, Astrophysical Institute Potsdam, University of Basel, University of Cambridge, Case Western Reserve University, University of Chicago, Drexel University, Fermilab, the Institute for Advanced Study, the Japan Participation Group, Johns Hopkins University, the Joint Institute for Nuclear Astrophysics, the Kavli Institute for Particle Astrophysics and Cosmology, the Korean Scientist Group, the Chinese Academy of Sciences (LAMOST), Los Alamos National Laboratory, Max-Planck-Institut f\"ur Astronomie (MPIA Heidelberg), Max-Planck-Institut f\"ur Astrophysik (MPA Garching), New Mexico State University, Ohio State University, University of Pittsburgh, University of Portsmouth, Princeton University, the United States Naval Observatory, and the University of Washington.

SDSS-IV is managed by the Astrophysical Research Consortium for the Participating Institutions of the SDSS Collaboration including the  Brazilian Participation Group, the Carnegie Institution for Science, Carnegie Mellon University, the Chilean Participation Group, the French Participation Group, Harvard-Smithsonian Center for Astrophysics, Instituto de Astrof\'isica de Canarias, The Johns Hopkins University, Kavli Institute for the Physics and Mathematics of the Universe (IPMU) / University of Tokyo, Lawrence Berkeley National Laboratory, Leibniz Institut f\"ur Astrophysik Potsdam (AIP),  Max-Planck-Institut f\"ur Astronomie (MPIA Heidelberg), Max-Planck-Institut f\"ur Astrophysik (MPA Garching), Max-Planck-Institut f\"ur Extraterrestrische Physik (MPE), National Astronomical Observatory of China, New Mexico State University, New York University, University of Notre Dame, Observat\'ario Nacional / MCTI, The Ohio State University, Pennsylvania State University, Shanghai Astronomical Observatory, United Kingdom Participation Group, Universidad Nacional Aut\'onoma de M\'exico, University of Arizona, University of Colorado Boulder, University of Oxford, University of Portsmouth, University of Utah, University of Virginia, University of Washington, University of Wisconsin, Vanderbilt University, and Yale University.

\textit{All data taken as part of SDSS-IV is scheduled to be released to the community in fully reduced form at regular intervals through dedicated data releases. The first MaNGA data release will be part of the SDSS data release 13 (release date 31 July 2016) and will be presented in a future paper from the SDSS collaboration.}
\end{small}


\bibliography{bib12}
\bibliographystyle{mnras}


\noindent \hrulefill

\noindent $^1$ University of Cambridge, Cavendish Astrophysics, Cambridge, CB3 0HE, UK.
\\$^2$ University of Cambridge, Kavli Institute for Cosmology, Cambridge, CB3 0HE, UK.
\\$^{3}$ Institute of Cosmology and Gravitation, University of Portsmouth, Dennis Sciama Building, Portsmouth, PO1 3FX, UK.
\\$^4$ European Southern Observatory, Karl-Schwarzchild-str., 2, Garching b. Munchen, 85748, Germany.
\\$^{5}$ Universit\'e Lyon 1, Observatoire de Lyon, Centre de Recherche Astrophysique de Lyon and Ecole Normale
Sup\'erieure de Lyon, 9 avenue Charles Andr\'e, Saint-Genis Laval, F-69230, France.
\\$^6$ University of Winsconsin-Madison, Department of Astronomy, 475 N. Charter Street, Madison, WI 53706-1582, USA.
\\$^{7}$ South East Physics Network (SEPNet), www.sepnet.ac.uk
\\$^8$ Department of Physics and Astronomy, University of Kentucky, 505 Rose Street, Lexington, KY 40506-0055, USA.
\\$^{9}$ Apache Point Observatory and New Mexico State University,, P.O. Box 59, Sunspot, NM 88349-0059, USA.
\\$^{10}$ Sternberg Astronomical Institute, Moscow State University, Moscow, Russia.
\\$^{11}$ Universidad de Antofagasta, Unidad de Astronomía, Avenida Angamos 601, Antofagasta, 1270300, Chile.
\\$^{12}$ Department of Physics and Astronomy, University of Utah, 115 S. 1400 E., Salt Lake City, UT 84112, USA.
\\$^{13}$ Kavli Institute for the Physics and Mathematics of the Universe (WPI), The University of Tokyo Institutes for Advanced Study, The University of Tokyo, Kashiwa, Chiba 277-8583, Japan.
\\$^{14}$ McDonald Observatory, The University of Texas at Austin, 2515 Speedway, Stop C1402, Austin, TX 78712, USA.
\\$^{15}$ Center for Astrophysical Sciences, Department of Physics and Astronomy, The Johns Hopkins University, Baltimore, MD 21218, USA.
\\$^{16}$ Space Telescope Science Institute, 3700 San Martin Drive, Baltimore, MD 21218, USA.
\\$^{17}$ Departamento de Fisica, Facultad de Ciencias, Universidad de La Serena, Cisternas 1200, La Serena, Chile.
\\$^{18}$ National Optical Astronomy Observatory, Tucson, AZ 85719, USA.
\\$^{19}$ School of Physics and Astronomy, University of St. Andrews, North Haugh, St. Andrews, KY16 9SS, UK.


\appendix

\section{Are galaxies really line-less?}
\label{appA}

Throughout the text we have defined `line-less' galaxies as galaxies with mean EW(H$\alpha$) $<$ 1.0 \AA\ within 1.0 $\rm R_e$. The chosen EW value is motivated by the wish to detect all the strong lines needed for BPT classification with adequate S/N at the typical depth of the MaNGA data and does not represent a physical boundary. Moreover, since we impose a cut on H$\alpha$, other lines might still be detectable in what we defined `line-less' galaxies, especially [NII], [OIII] and [SII], which are enhanced with respect to Balmer lines in the LIER regime. The study of the detection statistics of these lines in `line-less' galaxies is beyond the scope of this section.

In this work we have demonstrated that eLIER galaxies are characterised by low EW(H$\alpha$) ($<$ 3 \AA) and have equivalent stellar populations to line-less galaxies. If we make the assumption that line-less galaxies represent the continuation to lower EW of the eLIER population, it is interesting to explore the consequences for the EW(H$\alpha$) distribution for the combined eLIER and line-less classes.

Deriving EW(H$\alpha$) for line-less galaxies is a non-trivial exercise, as the continuum dominates the line emission and an accurate fit to the Balmer absorption is thus necessary. Moreover, at low flux level, the adopted Gaussian fitting strategy to recover emission line fluxes becomes biased, as the line shape becomes poorly defined. In order to attempt a measurement of the residual line emission in line-less galaxies we therefore perform the following: 1) The H$\alpha$ flux is calculated using the same binning scheme as the stellar continuum, in order to get the most accurate continuum subtraction possible, 2) The H$\alpha$ flux is calculated by performing a straight sum of the flux of the continuum-subtracted spectrum in a 600 $\rm km~s^{-1}$ window around the expected position of the H$\alpha$ line, taking the redshift of the source into account, and when detected, the measured velocity of the line. We refer to the line fluxes obtained this way as the `non-parametric' fluxes. We have checked that at high flux levels the line fluxes obtained from Gaussian fitting agree with the non-parametric fluxes.

\begin{figure} 
\includegraphics[width=0.5\textwidth, trim=80 90 80 150, clip]{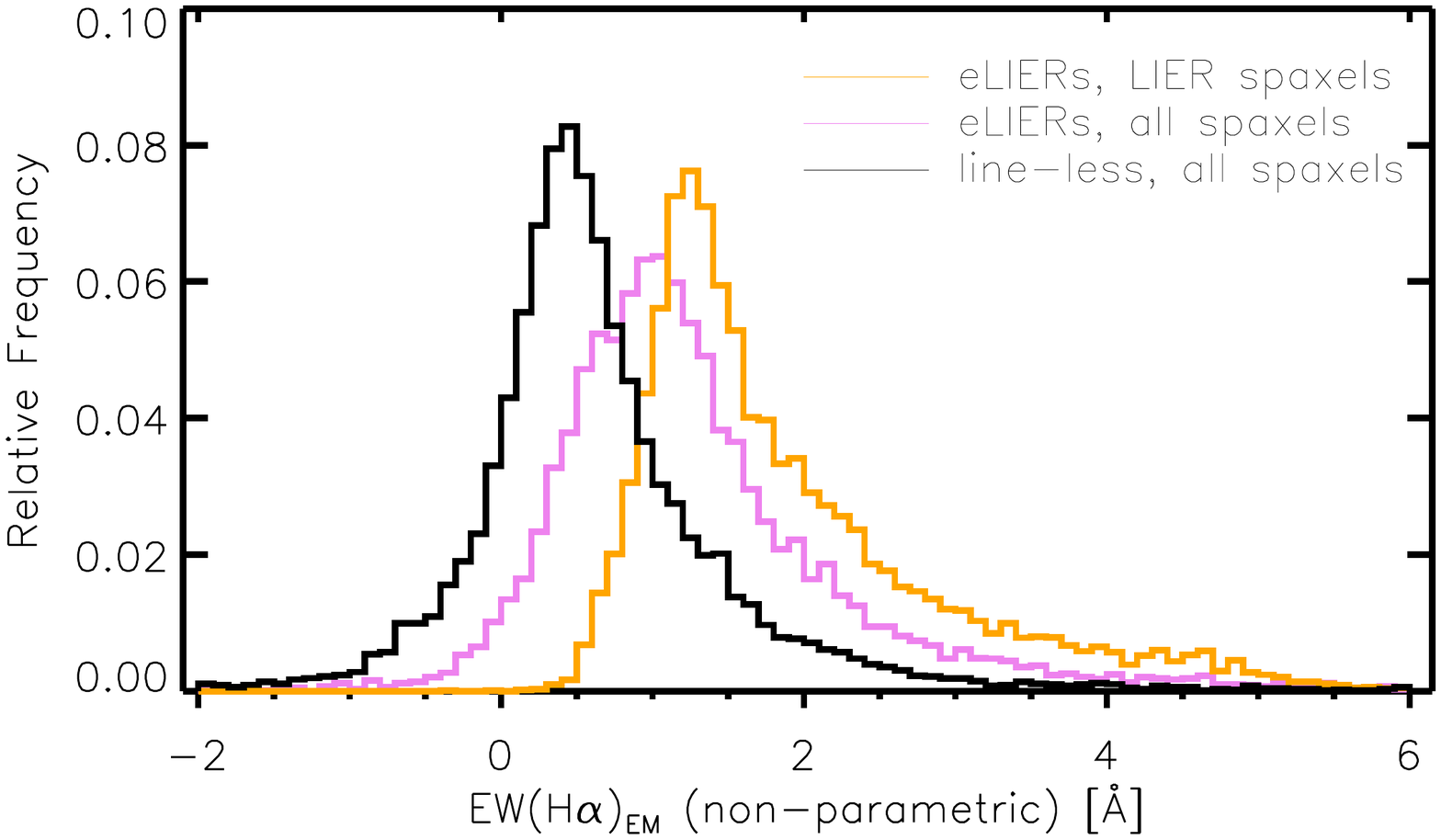}
\caption{Histogram of the distribution of EW(H$\alpha$) (in emission) calculated using the non-parametric method. Orange denotes regions with S/N $>$ 2 on all the [SII] BPT lines in eLIER galaxies (which are hence classified as LIERs in the [SII] BPT diagram). Violet corresponds to all regions within eLIER galaxies, and black to all regions in line-less galaxies.}
\label{fig_app3}
\end{figure}

In Fig. \ref{fig_app3} we show the EW(H$\alpha$) distribution obtained using non-parametric fluxes for the eLIER and line-less galaxies. Including spaxels with S/N $<$ 3 on the BPT strong lines lowers the median EW(H$\alpha$) for eLIER galaxies from 1.6 \AA\ to 1.1 \AA. The median EW(H$\alpha$) for line-less galaxies is 0.5 \AA. 

Further work, including deeper observations or stacking of the available MaNGA data, will be useful to decisively confirm the presence of residual line emission in `line-less' galaxies. The findings in this sections suggest that imposing a S/N cut sufficient to reliably measure all the BPT lines excludes a tail of lower EW(H$\alpha$) regions. The overall distribution of EW(H$\alpha$) for LIERs is therefore likely to be biased high because of the adopted S/N cut.

\section{Examples of eLIER and cLIER galaxies}
\label{appb}

The following figures are meant to illustrate a representative subset of cLIER and eLIER galaxies. For each galaxy we show:
\begin{enumerate}
\item{The $g-r-i$ SDSS colour composite image.}
\item{A map of the H$\alpha$ flux.}
\item{The position of the regions within the galaxy in the [SII] BPT diagram.}
\item{A map of classification of the galactic regions according to the [SII] BPT diagram. Blue represents SF regions, orange LIER-like and red Sy-like.}
\end{enumerate}
In all maps the MaNGA PSF is shown in the bottom right hand corner. The grey area corresponds to the size of the MaNGA bundle.

\begin{figure*} 
\includegraphics[width=1.0\textwidth, trim=0 80 10 60, clip]{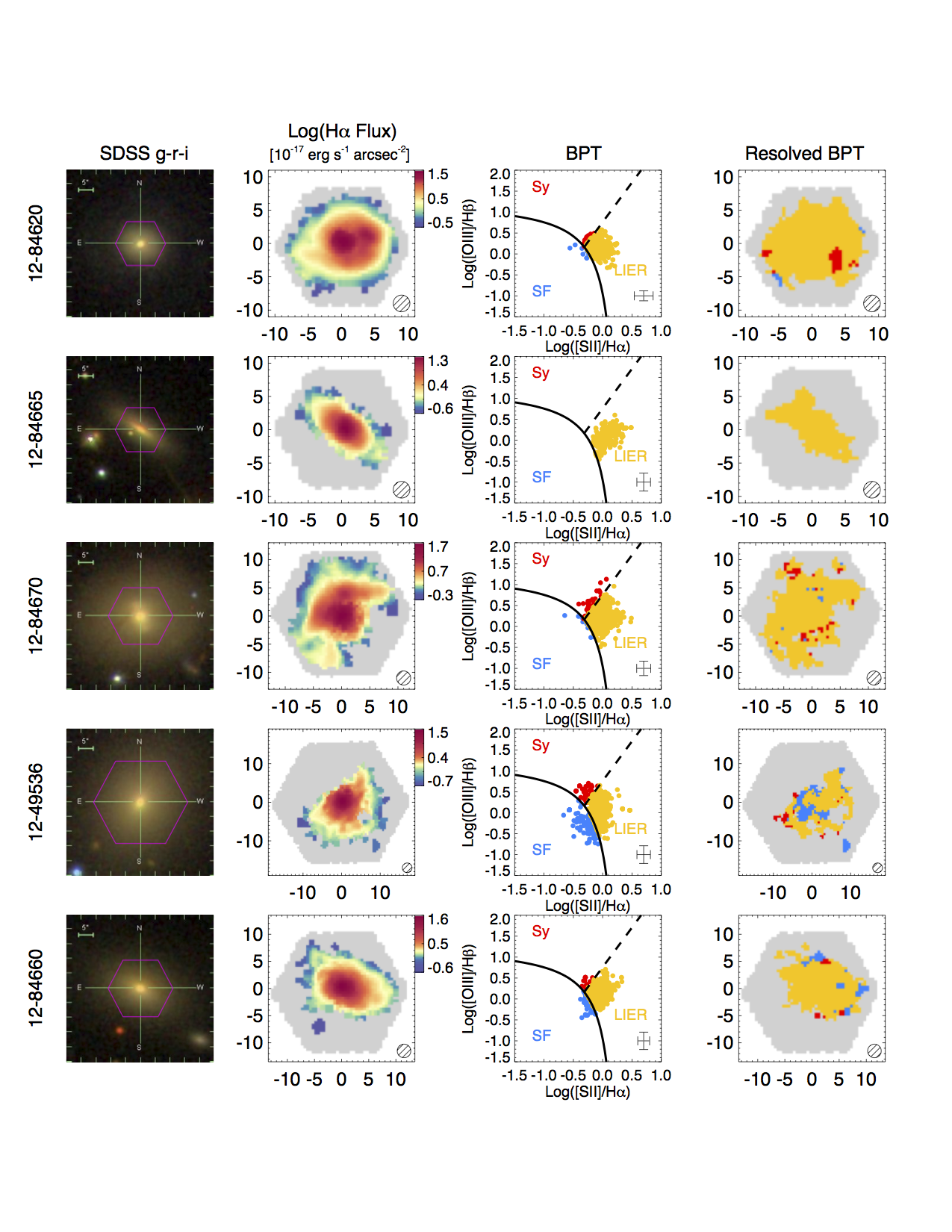}
\caption{Examples of extended LIER (eLIER) galaxies.}
\label{fig_ex1}
\end{figure*}

\begin{figure*} 
\includegraphics[width=1.0\textwidth, trim=0 80 10 60, clip]{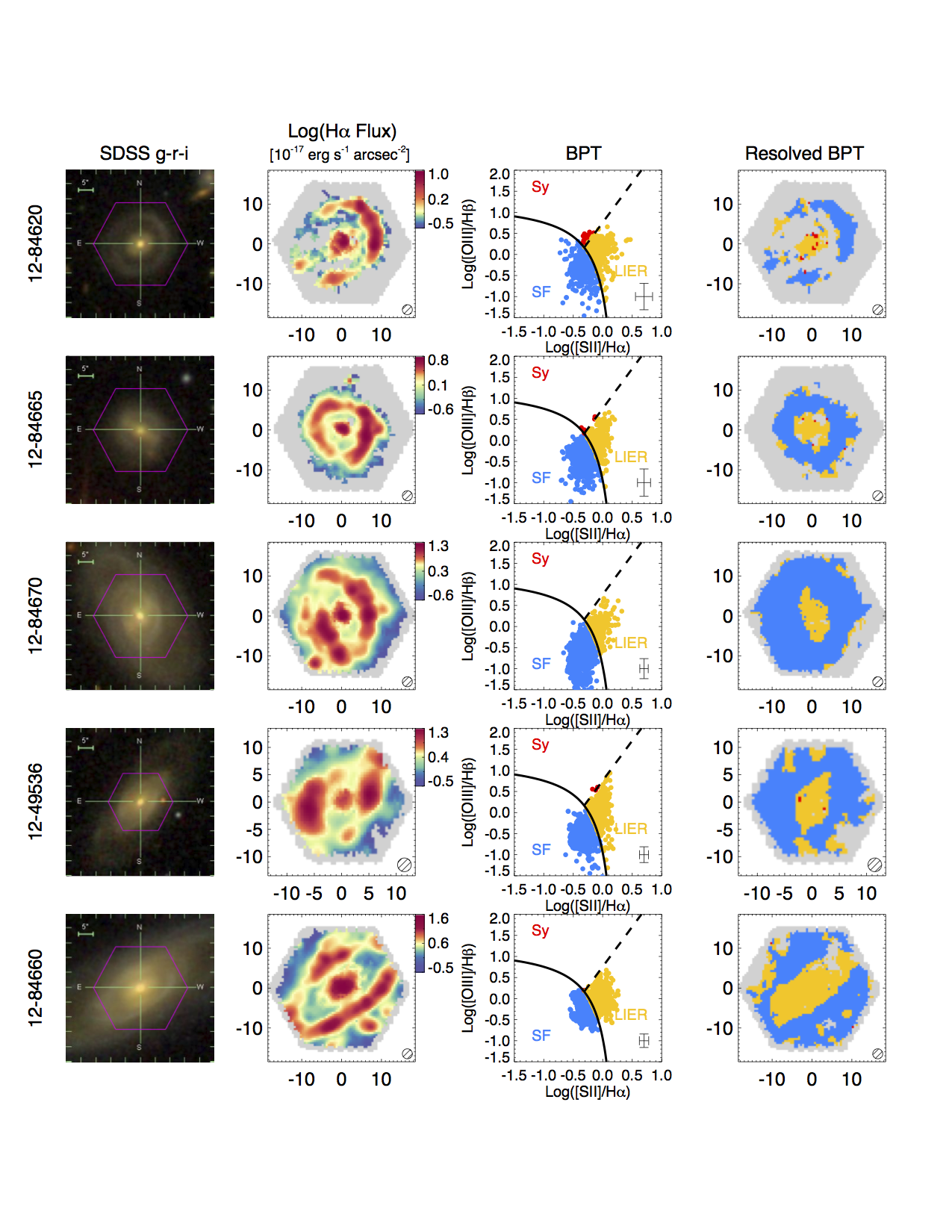}
\caption{Examples of central LIER (cLIER) galaxies.}
\label{fig_ex2}
\end{figure*}

\bsp
\label{lastpage}
\end{document}